\documentclass[pdflatex,sn-basic]{jnl}
\usepackage{graphicx}%
\usepackage{multirow}%
\usepackage{amsmath,amssymb,amsfonts}%
\usepackage{amsthm}%
\usepackage{mathrsfs}%
\usepackage[title]{appendix}%
\usepackage{xcolor}%
\usepackage{textcomp}%
\usepackage{manyfoot}%
\usepackage{booktabs}%
\usepackage{enumitem}
\usepackage{algorithm}%
\usepackage{algorithmicx}%
\usepackage{algpseudocode}%
\usepackage{listings}%
\theoremstyle{thmstyleone}%

\usepackage{siunitx}
\theoremstyle{thmstyletwo}%

\theoremstyle{thmstylethree}%

\begin{document}

\title[ME-Halo Orbits]{High-order expansions of multi-revolution elliptic Halo orbits in the elliptic restricted three-body problem}

\author[1,2]{\fnm{Xiaoyan} \sur{Leng}}
\author*[1,2]{\fnm{Hanlun} \sur{Lei}}\email{leihl@nju.edu.cn}



\affil[1]{\orgdiv{School of Astronomy and Space Science}, \orgname{Nanjing University}, \orgaddress{\city{Nanjing}, \postcode{210023}, \country{China}}}
\affil[2]{\orgdiv{Key Laboratory of Modern Astronomy and Astrophysics in Ministry of Education}, \orgname{Nanjing University}, \orgaddress{\city{Nanjing}, \postcode{210023}, \country{China}}}




\abstract{Multi-revolution elliptic Halo (ME-Halo) orbits are a special class of symmetric and periodic solutions within the framework of the elliptic restricted three-body problem (ERTBP). During a single period, an M:N ME-Halo orbit completes $M$ revolutions around a libration point and the primaries revolve $N$ times around each other. Owing to the repeated configurations, ME-Halo orbits hold great promise as nominal trajectories for space mission design. However, a major challenge associated with ME-Halo orbits lies in their mathematical description. To this end, we propose a novel method to derive high-order analytical expansions of ME-Halo orbits in the ERTBP by introducing two correction terms into the equations of motion in the $y$- and $z$-directions. Specifically, both the coordinate variables and correction terms are expanded as power series in terms of the primary eccentricity, the in-plane amplitude, and the out-of-plane amplitude. High-order approximations are constructed using a perturbation method, and their accuracy is validated through numerical analysis. Due to the inherent symmetry, ME-Halo orbits can be classified into four distinct families: southern/northern and periapsis/apoapsis groups. The analytical approximations developed in this study not only provide high-accuracy initial guesses for the numerical computation of ME-Halo orbits, but also offer new insights into the dynamical environment near collinear libration points in the ERTBP, thereby advancing practical applications in mission design.}

\keywords{Perturbation method, Multi-revolution elliptic Halo orbits, Double resonance}

\maketitle

\section{Introduction}\label{sec1}

The three-body problem remains one of the most enduring and fundamental challenges in celestial mechanics. In particular, in the circular restricted three-body problem (CRTBP) there are five equilibrium points, at which the gravitational and centrifugal forces are precisely balanced. These points are of particular interest for space missions \citep{bernelli2004assessment,masdemont2005high,peng2015low,neelakantan2022two,chujo2024quasi,burattini2024aerogel,conti2025design,sanaga2025leveraging,gao2025bounds,yamaguchi2025trajectory}. 

Regarding the issue of CRTBP, \citet{szebehely1967} provided a comprehensive review about analytical and numerical methods to investigate the dynamics near libration points. Following this methodological framework, \citet{farquhar1973} employed the Lindstedt--Poincar\'e (L--P) method to explore quasi-periodic orbits near the Earth–Moon libration points, successfully revealing the existence of Halo orbits. \citet{richardson1980analytic} further applied the L--P method to derive a third-order analytical solution of Halo orbits, which has been widely used to provide initial guesses for computing Halo orbits. 
Based on partial normalization via the Lie series method, \citet{jorba1999} reduced the dynamics near the equilibrium points to the center manifold, where the dynamical behavior was typically analyzed by means of Poincar\'e sections. The L--P technique was adopted to formulate high-order expansions of center manifolds, including Lissajous and Halo orbits. This topic was extended by \citet{masdemont2005high} to invariant manifolds, enabling accurate modeling of motion near libration points and supporting practical applications such as low-energy transfers, station-keeping, and trajectory design between $L_1$ and $L_2$. 
In addition, it shows that the Koopman Operator can be used to formulate analytical solutions for Lyapunov and Halo orbits \citep{servadio2023koopman}. 

Analytical progress has been made in understanding the bifurcation mechanisms of Halo orbits. 
By constructing a normal form adapted to the synchronous resonance, and introducing a detuning parameter to characterize the frequency deviation, \citet{ceccaroni2016halo} derived energy thresholds for the onset of Halo orbit bifurcations under arbitrary mass ratios.
\citet{luo2020lissajous} obtained analytical forms of normalized Hamiltonian functions, and derived the formula for the bifurcation energy. Furthermore, they provided the analytical solutions of Lissajous and halo orbits through an inverse transformation. Recently, \citet{lin2024Bifurcation} proposed a generalized dynamical framework by incorporating a coupling coefficient, which facilitates the unified description of Lissajous, Halo, and quasi-Halo orbits.

While the CRTBP remains applicable in certain contexts, its limitations are well recognized. In particular, \citet{parker2014} identified that the non-zero eccentricity of the primaries’ orbits may lead to deviations from orbital periodicity. To this end, the elliptic restricted three-body problem (ERTBP) provides a more accurate representation of spacecraft dynamics. Within the framework of ERTBP, \citet{heppenheimer1973} investigated the out-of-plane motion near the libration points and analyzed the influence of amplitude and eccentricity upon the orbital periods. From an analytical viewpoint, \citet{hou2011} developed high-order solutions for Lissajous and Halo orbits, and classified three types of symmetric periodic orbits. About the same topic, \citet{lei2013} extended to high-order expansions of invariant manifolds around collinear libration points in the ERTBP. \citet{paez2022semi} advanced the theory of Halo orbits in the ERTBP by developing high-order approximations of quasi-periodic orbits and their associated invariant manifolds, employing a resonant Floquet--Birkhoff normal form. Recently, \citet{Celletti2024Elliptic} applied the normalization technique to the ERTBP, providing analytical solutions for Lyapunov and Halo orbits near the libration points, along with expressions for the bifurcation energies.

Due to the periodic perturbations introduced by the orbital eccentricity of the secondary body, periodic orbits in the CRTBP generally lose strict periodicity when extended to the ERTBP, evolving into quasi-periodic trajectories. Nevertheless, strictly periodic orbits do exist within the framework of ERTBP. As early as 1920, \citet{moulton1920} proposed a geometric criterion for strong periodicity in the ERTBP. Later, \citet{sarris1989} demonstrated that, unlike in the CRTBP—where orbital periods can assume arbitrary values—only integer multiples of $2\pi$ are permitted in the ERTBP. Building upon these foundational insights, \citet{campagnola2008} coined the term `elliptic Halo orbits' to describe the ERTBP counterparts of CRTBP Halo orbits and computed them using eccentricity-based numerical continuation with CRTBP Halo orbits as initial guesses. \citet{peng2015} further referred to these trajectories as `multi-revolution elliptic Halo (ME-Halo) orbits'\footnote{We adopt the terminology `multi-revolution elliptic Halo orbits' (ME-Halo orbits for short) in this study.} and employed the technique of arc-length continuation to avoid singularities associated with single-parameter continuation methods. Using the variation of constants, \citet{Asano2015Approximating} formulated a second-order approximation for ME-Halo orbits around collinear libration points. In a systematic study, \citet{ferrari2018periodic} extended Lyapunov, vertical Lyapunov, and Halo orbits from the CRTBP to the ERTBP, obtaining single-, double-, triple-, and quadruple-revolution Halo orbits. \citet{neelakantan2021design} introduced a differential evolution-based method for designing multi-revolution periodic orbits, enabling the computation of multi-revolution Lyapunov and Halo orbits. For solar sails in the Sun–Earth ERTBP, \citet{huang2020families} constructed ME-Halo orbits and identified bifurcations based on stability analysis. Periodic orbits have been investigated in \citet{zheng2024universal} by numerically continuing models from the CRTBP to the ERTBP. Applications of ME-Halo orbits in mission design have been widely discussed \citep{peng2015low,peng2015transfer,peng2017maintenance,peng2018natural,neelakantan2022two}.

From the aforementioned studies, we can see that numerical methods have thus far been the primary means for generating ME-Halo orbits within the framework of ERTBP. 
However, a key challenge of these approaches lies in their sensitivity to initial conditions. Initial guesses derived from Halo orbits in the CRTBP typically require successive corrections through eccentricity continuation and multi-shooting methods, thereby increasing the computational complexity\citep{campagnola2008,peng2015,ferrari2018periodic}.
This raises a fundamental question: is it possible to develop an analytical approximation for ME-Halo orbits directly within the ERTBP framework? The primary objective of this study is to answer this question. We hope to construct an analytical approximation that not only provides high-accuracy initial guesses for the numerical construction of ME-Halo orbits but also offers new insights into the dynamical behavior near collinear libration points, thereby advancing both theoretical understanding and practical applications.    

The remainder of this paper is organized as follows. In Section \ref{sec2}, we briefly introduce the dynamical model. Section \ref{sec3} shows the details of formulating high-order expansions of ME-Halo orbits in the framework of ERTBP. Analytical results of ME-Halo orbits are given in Section \ref{sec4}, and numerical explorations are made in Section \ref{sec5}. At last, the main conclusion is summarized in Section \ref{sec6}.

\section{Elliptic restricted three-body problem}
\label{sec2}

The elliptic restricted three-body problem (ERTBP) studies the motion of a massless third body, denoted by $P_3$, under the gravitational field of two primaries, denoted by $P_1$ and $P_2$. Let us denote the mass of the third body as $m$, the masses of the primary bodies as $m_1$ and $m_2$. Without loss of generality, we assume $m_1 > m_2$. In the test-particle limit, it holds $m \ll m_2 < m_1$, showing that $P_3$ has negligible influence on the motion of the primaries. Thus, the primaries move around their barycenter in Keplerian orbits with eccentricity of $e$. In particular, when the eccentricity is $e=0$, the ERTBP reduces to the circular restricted three-body problem (CRTBP). 

Usually, the motion of $P_3$ is described in a barycentric synodic coordinate system, where the $X$-axis points from $P_1$ toward $P_2$, the $Z$-axis is aligned with the angular momentum vector of the primaries, and the $Y$-axis is chosen to complete the right-handed rule. For convenience, the following system of units is used: the instantaneous distance between the primaries is taken as the unit of length, the total mass of primaries as the unit of mass, and the orbital period divided by $2\pi$ as the time unit. Under the normalized system of units, both the mean motion of the primaries and the universal gravitational constant become unity. In addition, the secondary $P_2$ with normalized mass of $\mu = m_2 / (m_1 + m_2)$ is located at $(1-\mu,0,0)$, and the primary $P_1$ with normalized mass of $1-\mu$ is located at $(-\mu,0,0)$. 

In the ERTBP, the equations of motion for the test particle $P_3$ can be written as \citep{szebehely1967}
\begin{equation}
X'' - 2 Y' = \frac{1}{1 + e \cos f} \frac{\partial{\Omega}}{\partial{X}},\; Y'' + 2 X' = \frac{1}{1 + e \cos f} \frac{\partial{\Omega}}{\partial{Y}},\; Z'' = \frac{1}{1 + e \cos f} \frac{\partial{\Omega}}{\partial{Z}},
\label{EQ1}
\end{equation}
where the effective potential is given by
\begin{equation}
    \Omega =\frac{1}{2}\left[{X^{2}}+{Y^{2}}+{Z^{2}}+\mu (1-\mu )\right]+\frac{1-\mu }{{{R}_{1}}}+\frac{\mu }{{{R}_{2}}}  
\end{equation}
with $R_1$ and $R_2$ being the distances of $P_3$ from $P_1$ and $P_2$, respectively. In Eq. \eqref{EQ1}, $f$ is the true anomaly of the primary orbit and the prime symbol denotes the derivative with respect to the true anomaly. With the $X$ component as an example, it reads
\begin{equation}
    X'=\frac{{\rm d}X}{{\rm d}f}, \quad X''=\frac{{\rm d}^2X}{{\rm d}f^2}.
\end{equation}
In the ERTBP, there are five equilibrium points \citep{szebehely1967}, three of them are located on the $X$-axis, called collinear libration points, and the remaining two constitute equilateral triangles with the primaries, called triangular libration points. The focus of this work is to study the motion around collinear libration points. To this end, it is required to first move the origin of coordinate system to the libration point of interest, and then take the instantaneous distance between the considered libration point and its nearest primary $\gamma_i$ as the new unit of length. The coordinate transformation between the original barycentric synodic frame and the $L_i$-centered frame can be realized by
\begin{equation}
    X={{\gamma }_{i}}\left( x\mp 1 \right)+1-\mu ,\ Y={{\gamma }_{i}}y,\ Z={{\gamma }_{i}}z, \quad i=1,2
\end{equation}
where the upper sign refers to the $L_1$ point and the lower one refers to the $L_2$ point ($L_3$ is not considered in this study). In the $L_i$-centered synodic system, the equations of motion can be organized as follows \citep{hou2011,lei2013}:
\begin{equation}
\begin{aligned}
   &{x}''-2{y}'=
   \sum\limits_{i\ge 0}{{{(-e)}^{i}}{{\cos }^{i}}f(1+2{{c}_{2}})x}+\sum\limits_{i\ge 0}{\sum\limits_{n\ge 2}{  {{(-e)}^{i}}{{\cos }^{i}}f {{c}_{n+1}}(n+1) {{T}_{n}}}},\\
   &{y}''+2{x}'=
   \sum\limits_{i\ge 0}{{{(-e)}^{i}}{{\cos }^{i}}f(1-{{c}_{2}})y}+\sum\limits_{i\ge 0}{  \sum\limits_{n\ge 2}{ {{(-e)}^{i}}{{\cos }^{i}}f {{c}_{n+1}} y {{R}_{n-1}}}},  \\
   &{z}''+{{c}_{2}}z=
   \sum\limits_{i\ge 1}{{{(-e)}^{i}}{{\cos }^{i}}f(1-{{c}_{2}})z}+\sum\limits_{i\ge 0}{ \sum\limits_{n\ge 2}{ {{(-e)}^{i}}{{\cos }^{i}}f {{c}_{n+1}} z {{R}_{n-1}}}}.
\end{aligned}
\label{eqfinal}
\end{equation}
In the above equations, $c_n$ is determined by the mass parameter $\mu$, given by
\begin{equation}
{{c}_{n}}=\frac{1}{\gamma _{i}^{3}}[{{(\pm 1)}^{n}}\mu +{{(-1)}^{n}}\frac{(1-\mu )\gamma _{i}^{n+1}}{{{(1\mp {{\gamma }_{i}})}^{n+1}}} ],\; i=1,2
\end{equation}
where the upper sign is for the $L_1$ point and the lower sign is for the $L_2$ point. Additionally, $T_n$ and $R_{n-1}$ are defined by \citep{jorba1999}
\begin{equation}
{{T}_{n}}={{\rho }^{n}}{{P}_{n}}(\frac{x}{\rho } ), \quad 
    {{R}_{n-1}}=\frac{1}{y}\frac{\partial {{T}_{n+1}}}{\partial y}=\frac{1}{z}\frac{\partial {{T}_{n+1}}}{\partial z}, 
\end{equation}
where $P_n(\cdot)$ is the Legendre polynomial and $\rho$ is the distance $\rho = \sqrt{x^2 + y^2 + z^2}$. In particular, $T_n$ and $R_n$ are homogeneous polynomials of degree $n$ of coordinates, computed recursively by \citep{jorba1999}
\begin{equation}
\begin{aligned}
    {{T}_{n}}&=\frac{2n-1}{n}x{{T}_{n-1}}-\frac{n-1}{n}({{x}^{2}}+{{y}^{2}}+{{z}^{2}}){{T}_{n-2}}, \\ 
    {{R}_{n}}&=\frac{2n+3}{n+2}x{{R}_{n-1}}-\frac{2n+2}{n+2}{{T}_{n}}-\frac{n+1}{n+2}({{x}^{2}}+{{y}^{2}}+{{z}^{2}}){{R}_{n-2}},
\end{aligned}
\end{equation} 
starting with $T_0 = 1$, $T_1 = x$, $R_0 = -1$, and $R_1 = -3x$.

\section{Analytic construction of ME-Halo orbits}
\label{sec3}

In the ERTBP, the perturbation introduced by the orbital eccentricity $e$ typically causes Halo orbits to exhibit quasi-periodic behavior. However, when the motion frequency of a Halo orbit forms a simple integer ratio with the system’s forced frequency, resonance effects can make the orbit become periodic. According to the definition given by \citet{peng2015}, a periodic orbit is defined as the $M$:$N$ ME-Halo orbit if its period $T_E$ is an integer multiple of the system’s fundamental period $2\pi$ ($T_E = N \cdot 2\pi$), and also an integer multiple of the corresponding Halo orbit period $T_C$ in the CRTBP ($T_E = M \cdot T_C$)\footnote{The period of CRTBP Halo orbit is $T_C=2\pi/\omega=2\pi/\nu$, where $\omega$ and $\nu$ are the in-plane and out-of-plane motion frequencies, respectively.}. Without loss of generality, in this section we take M2N1 ME-Halo orbits as an example to discuss the construction of high-order series expansions.

\subsection{The basic idea}
\label{subsec31}

It is known that Halo orbits in the CRTBP are a type of three-dimension periodic orbits, which are bifurcated from the family of planar Lyapunov orbits. Thus, the emergence of Halo orbits is due to the frequency degeneracy condition, where the frequency of in-plane motion $\omega$ is exactly equal to that of out-of-plane motion $\nu$. To realize the Halo orbit condition of $\omega=\nu$ in the CRTBP, \citet{richardson1980analytic} and \citet{jorba1999} introduced a correction term $\Delta z$ in the $z$-direction equation of motion and developed high-order expansions of Halo orbits by means of L--P method. Similarly, \citet{hou2011} and \citet{lei2013} introduced the correction term $\Delta z$ in the $z$-direction equation of motion to reach the condition of $\omega=\nu$ in the ERTBP. Particularly, here $\Delta$ must also be expanded as a power series of in-plane and out-of-plane amplitudes, where the coefficients are solved iteratively via the L--P method. The correction condition of $\Delta =0$ indicates that in-plane and out-of-plane amplitudes (usually denoted by $\alpha$ and $\beta$) of Halo orbits are no longer independent. Thus, we can see that the degeneracy condition of fundamental frequencies ($\omega=\nu$) is realized by introduction of $\Delta=0$. 

Generally, Halo orbits with $\omega=\nu$ in the ERTBP are no longer periodic due to the existence of an additional forced frequency caused by the primary's eccentricity. Only when the forced frequency and the in-plane (or out-of-plane) frequency are still commensurable, the Halo orbits may become periodic. The resulting periodic Halo orbits are termed ME-Halo orbits. Thus, we can interpret ME-Halo orbits as a class of orbits embedded within double resonance structures. In particular, one resonance involves a commensurability between two natural frequencies, whereas the other involves a commensurability between a natural and a forced frequency. 

Inspired by the construction of Halo orbits in the CRTBP, we propose a high-order expansion approach for ME-Halo orbits by introducing two correction terms: (a) the first correction term, denoted as $\Delta_1 y$, is added to the $y$-direction equation of motion\footnote{It is worth noting that applying the same method to introduce $\Delta_1 x$ for correction in the $x$-direction yields parameter values that deviate significantly from those obtained through correction in the $y$-direction. Moreover, when the analytical solution derived from the $x$-direction correction is used as the initial condition for numerical integration, the resulting trajectory exhibits divergence. The underlying cause of this pronounced discrepancy in the $x$-direction correction remains unclear and warrants further investigation.} to ensure that the forced frequency is commensurable with either the in-plane or out-of-plane natural frequency; and (b) the second correction term, denoted as $\Delta_2 z$, is introduced in the $z$-direction equation of motion to enforce a 1:1 resonance between the in-plane and out-of-plane motions ($\omega = \nu$).

The equations of motion with correction terms are written as
\begin{equation}
\begin{aligned}
&{x}''-2{y}'=
\sum\limits_{i\ge 0}{{{(-e)}^{i}}{{\cos }^{i}}f(1+2{{c}_{2}})x}+\sum\limits_{i\ge 0}{\sum\limits_{n\ge 2}{  {{(-e)}^{i}}{{\cos }^{i}}f {{c}_{n+1}}(n+1) {{T}_{n}}}},\\
&{y}''+2{x}'=
\sum\limits_{i\ge 0}{{{(-e)}^{i}}{{\cos }^{i}}f(1-{{c}_{2}})y}+\sum\limits_{i\ge 0}{  \sum\limits_{n\ge 2}{ {{(-e)}^{i}}{{\cos }^{i}}f {{c}_{n+1}} y {{R}_{n-1}}}} + \Delta_1 y,  \\
&{z}''+{{c}_{2}}z=
\sum\limits_{i\ge 1}{{{(-e)}^{i}}{{\cos }^{i}}f(1-{{c}_{2}})z}+\sum\limits_{i\ge 0}{ \sum\limits_{n\ge 2}{ {{(-e)}^{i}}{{\cos }^{i}}f {{c}_{n+1}} z {{R}_{n-1}}}} + \Delta_2 z.
\end{aligned}
\label{eq21Halo}
\end{equation}
The invariance of equations of motion requires the correction conditions of $\Delta_1=0$ and $\Delta_2=0$, leading to implicit relationships among the eccentricity $e$, in-plane amplitude $\alpha$ and out-of-plane amplitude $\beta$.

\subsection{Third-order analytical solution}
\label{subsec32}

Perturbation method is employed to derive third-order analytical solutions for ME-Halo orbits through a recursive process, starting from the linear solution \citep{Lei2024perturbation}. It should be noted that the third-order solution represents the lowest order at which nonlinear corrections associated with ME–Halo resonances emerge, just like the third-order solution of Halo orbits in the CRTBP \citep{richardson1980analytic}.
To formulate third-order solution, the equations of motion are truncated up to third order with respect to the variables $(e, x, y, z)$ as follows:
\begin{equation}
\begin{aligned}
   {x}''-2{y}'-&\left( 1+2{{c}_{2}} \right)x=
  \frac{3}{2}{{c}_{3}}\left( 2{{x}^{2}}-{{y}^{2}}-{{z}^{2}} \right)-\left( 1+2{{c}_{2}} \right)ex\cos f \\
  &+\frac{1}{2}\left( 1+2{{c}_{2}} \right)\left( 1+\cos 2f \right){{e}^{2}}x-\frac{3}{2}{{c}_{3}}\left( 2e{{x}^{2}}-e{{y}^{2}}-e{{z}^{2}}\right)\cos f\\
  &+2{{c}_{4}}x\left( 2{{x}^{2}}-3{{y}^{2}}-3{{z}^{2}} \right),\\ 
  {y}''+2{x}'-&\left( 1-{{c}_{2}} \right)y=
 -3{{c}_{3}}xy-\left( 1-{{c}_{2}} \right)ey\cos f \\ 
 & +\frac{1}{2}\left( 1-{{c}_{2}} \right)\left( 1+\cos 2f \right){{e}^{2}}y+3{{c}_{3}}exy\cos f\\
 &-\frac{3}{2}{{c}_{4}}y\left( 4{{x}^{2}}-{{y}^{2}}-{{z}^{2}} \right)+{{\Delta }_{1}}y,\\ 
  {z}''+{{c}_{2}}z&=
 -3{{c}_{3}}xz-\left( 1-{{c}_{2}} \right)ez\cos f+\frac{1}{2}\left( 1-{{c}_{2}} \right)\left( 1+\cos 2f \right){{e}^{2}}z\\
 &+3{{c}_{3}}exz\cos f-\frac{3}{2}{{c}_{4}}z\left( 4{{x}^{2}}-{{y}^{2}}-{{z}^{2}} \right)+{{\Delta }_{2}}z.  
\end{aligned}
\label{eq3rd}
\end{equation}
Up to the third order in terms of $(e, x, y, z)$, the coordinates $(x, y, z)$ as well as the correction terms ($\Delta_1$ and $\Delta_2$) are expanded as follows:
\begin{equation}
   \begin{aligned}
     & x={{x}_{1}}+{{x}_{2}}+{{x}_{3}}, \\ 
     & y={{y}_{1}}+{{y}_{2}}+{{y}_{3}}, \\ 
     & z={{z}_{1}}+{{z}_{2}}+{{z}_{3}}, \\
     & {\Delta _1}={a_0}+{a_1}+{a_2},\\ 
     & {\Delta _2}={b_0}+{b_1}+{b_2}.\\ 
    \end{aligned} 
    \label{eqxyzab}
\end{equation}
By replacing Eq. \eqref{eqxyzab} in Eq. \eqref{eq3rd}, it is possible for us to construct the analytical solution by taking advantage of the method of undetermined coefficients. 

Initially, the linearized equations of motion are
\begin{equation}
\begin{aligned}
&{{x_1}''}-2{{y_1}'}-\left( 1+2{{c}_{2}} \right){{x}_{1}}=0,\\ 
&{{y_1}''}+2{{x_1}'}-\left( 1-{{c}_{2}}+{{a}_{0}} \right){{y}_{1}}=0,\\ 
&{{z_1}''}+\left( {{c}_{2}}-{{b}_{0}} \right){{z}_{1}}=0.\\
\end{aligned}
\label{eq1}
\end{equation}
By solving Eq. \eqref{eq1}, it is not difficult to obtain the first-order solution of M2N1 ME-Halo orbit as follows\footnote{As for the M2N1 ME-Halo orbit, the in-plane and out-of-plane motion frequencies are fixed at $\omega=\nu=2n$, where the forced frequency is $n=1$ in normalized units.}:
\begin{equation}
    {{x}_{1}}=\alpha \cos 2f, \quad
    {{y}_{1}}=\kappa \alpha \sin 2f, \quad
    {{z}_{1}}=\beta \cos 2f, 
\label{jie1}
\end{equation}
where $\alpha$ and $\beta$ are referred to as the in-plane and out-of-plane amplitudes, and $\kappa$ is
\begin{equation}
    \kappa =-\frac{1}{4} \left(5+2{{c}_{2}}\right).
\end{equation}
Replacing Eq. \eqref{jie1} in Eq. \eqref{eq1}, we can obtain the zero-order coefficients of correction terms, given by
\begin{equation}
    {{a}_{0}}=-\frac{9+c_2(5-2c_{2})}{5+2{{c}_{2}}} , \quad {{b}_{0}}={{c}_{2}}-4.
\end{equation}
Notice that, for a certain system with given mass parameter $\mu$, the coefficients $a_0$, $b_0$, and $\kappa$ are constants. At the first order, we can get $\Delta_1 = a_0 \ne 0$ and $\Delta_2 = b_0 \ne 0$, showing that there is no linear solution for ME-Halo orbits\footnote{This is in agreement with Halo orbits in the CRTBP \citep{richardson1980analytic}.}.

Then, at the second order, the equations of motion are
\begin{equation}
    \begin{aligned}
         &{{x_2}''}-2{{y_2}'}-\left( 1+2{{c}_{2}} \right){{x}_{2}} ={{C}_{x0}^{(2)}}+ {{C}_{x1}^{(2)}} \cos f + {{C}_{x3}^{(2)}}\cos 3f+ {{C}_{x4}^{(2)}}\cos 4f, \\ 
         &{{y_2}''}+2{{x_2}'}-\left( 1-{{c}_{2}}+{{a}_{0}} \right){{y}_{2}} = {{S}_{y1}^{(2)}}\sin f +   {{a}_{1}}\kappa \alpha \sin 2f
         +{{S}_{y3}^{(2)}}\sin 3f+{{S}_{y4}^{(2)}}\sin 4f, \\ 
        &{{z_2}''}+\left( {{c}_{2}}-{{b}_{0}} \right){{z}_{2}}  = {{C}_{z0}^{(2)}}+ {{C}_{z1}^{(2)}}\cos f
        +{{b}_{1}}\beta \cos 2f 
        +{{C}_{z3}^{(2)}}\cos 3f
        +{{C}_{z4}^{(2)}}\cos 4f,\\ 
\end{aligned}
\label{eq2}
\end{equation}
where the expressions of $C_*^{(2)}$ and $S_*^{(2)}$ are provided by Eq. \eqref{A1} in the Appendix. The second-order analytical solution can be written as
\begin{equation}
\begin{aligned}
    &{{x}_{2}}={{x}_{20}}+{{x}_{21}}\cos f+{{x}_{22}}\cos 2f+{{x}_{23}}\cos 3f+{{x}_{24}}\cos 4f,\\
    &{{y}_{2}}={{y}_{20}}+{{y}_{21}}\sin f+{{y}_{22}}\sin 2f+{{y}_{23}}\sin 3f+{{y}_{24}}\sin 4f,\\
    &{{z}_{2}}={{z}_{20}}+{{z}_{21}}\cos f+{{z}_{22}}\cos 2f+{{z}_{23}}\cos 3f+{{z}_{24}}\cos 4f.\\
\end{aligned}
\label{jie2}
\end{equation}
Replacing Eq. \eqref{jie2} in Eq. \eqref{eq2}, we can determine the second-order coordinate coefficients ($x_2,y_2,z_2$) and the first-order coefficients of correction terms ($a_1,b_1$). It should be noted that when solving for $x_{22}$, $y_{22}$, and $a_1$, the number of unknowns exceeds the number of equations. Without loss of generality, this issue can be resolved by setting $x_{22} = 0$. Similar treatment can be found in \citet{richardson1980analytic}. As a result, 
the first-order coefficients of correction terms are
\begin{equation}
    {{a}_{1}}=0 , \quad {{b}_{1}}=0,\\
\end{equation}
and those second-order non-zero coefficients of coordinates are
\begin{equation}
    \begin{aligned}
        & {{x}_{20}}=-\frac{{{C}_{x0}^{(2)}}}{\left( 1+2{{c}_{2}} \right)},\quad {{x}_{21}}=-\frac{(6c_2-1){C}_{x1}^{(2)} + 2(5+2c_2){S}_{y1}^{(2)} }{6 (3 + 3 c_2 + 2 c_2^2)},\\ 
        &{{x}_{23}}=- \frac{(41+10c_2){C}_{x3}^{(2)} - 6(5+2c_2){S}_{y3}^{(2)} }{10 (23 + 11 c_2 + 2 c_2^2)},\quad {{x}_{24}}=-\frac{(19 + 6 c_2) {C}_{x4}^{(2)} - 2(5+2c_2){S}_{y4}^{(2)}}{3{\left( 9+2{{c}_{2}} \right)}^{2}},\\
        &{{y}_{21}}=\frac{(5 + 2 c_2 ){C}_{x1}^{(2)} -(5+7c_2+2c_2^2){S}_{y1}^{(2)}}{3\left( 3+3{{c}_{2}}+2c_{2}^{2} \right)},\\
        &{{y}_{23}}=-\frac{-3(5+2c_2){C}_{x3}^{(2)} + (25+15c_2+2c_2^2){S}_{y3}^{(2)}}{5\left( 23+11{{c}_{2}}+2c_{2}^{2} \right)},\\ 
        & {{y}_{24}}=\frac{-8(5+2c_2){C}_{x4}^{(2)} + (85+44 c_2 + 4 c_2^2) {S}_{y4}^{(2)}}{12{{\left( 9+2{{c}_{2}} \right)}^{2}}},\\
        & {{z}_{20}}=\frac{{C}_{z0}^{(2)}}{4},\quad {{z}_{21}}=\frac{{C}_{z1}^{(2)}}{3},\quad {{z}_{23}}=-\frac{{C}_{z3}^{(2)}}{5}, \quad {{z}_{24}}=-\frac{{C}_{z4}^{(2)}}{12}.
    \end{aligned}
\label{co2}
\end{equation}

At last, the equations of motion for the third-order terms are  
\begin{equation}
\begin{aligned}
          {x_3}''-2{y_3}'&-\left( 1+2{c}_{2} \right){x}_{3}={{C}_{x0}^{(3)}}+{{C}_{x1}^{(3)}}\cos f+{{C}_{x2}^{(3)}}\cos 2f+{{C}_{x3}^{(3)}}\cos 3f\\ 
         &+{{C}_{x4}^{(3)}}\cos 4f+{{C}_{x5}^{(3)}}\cos 5f+{{C}_{x6}^{(3)}}\cos 6f, \\
          {{y_3}''}+2{{x_3}'}&-\left( 1-{{c}_{2}} \right){{y}_{3}}-{{a}_{0}}{{y}_{3}}-{{a}_{2}}{{y}_{1}}={{S}_{y1}^{(3)}}\sin f+ ({{S}_{y2}^{(3)}}+{{a}_{2}}\kappa \alpha )\sin 2f\\ 
        & +{{S}_{y3}^{(3)}}\sin 3f+{{S}_{y4}^{(3)}}\sin 4f+{{S}_{y5}^{(3)}}\sin 5f+{{S}_{y6}^{(3)}}\sin 6f, \\
         {{z_3}''}+{{c}_{2}}{{z}_{3}}&-{{b}_{0}}{{z}_{3}}-{{b}_{2}}{{z}_{1}}={{C}_{z0}^{(3)}}+{{C}_{z1}^{(3)}}\cos f+({{C}_{z2}^{(3)}}+{{b}_{2}}\beta )\cos 2f\\ 
        &+{{C}_{z3}^{(3)}}\cos 3f+{{C}_{z4}^{(3)}}\cos 4f+{{C}_{z5}^{(3)}}\cos 5f+{{C}_{z6}^{(3)}}\cos 6f,
    \end{aligned}
    \label{eq3}
\end{equation}
where the expressions of $C_*^{(3)}$ and $S_*^{(3)}$ are provided by Eqs. \eqref{A2} and \eqref{A3} in the Appendix. Similarly, the third-order solution can be expressed as
\begin{equation}
    \begin{aligned}
    {{x}_{3}}=&{{x}_{30}}+{{x}_{31}}\cos f+{{x}_{32}}\cos 2f+{{x}_{33}}\cos 3f+{{x}_{34}}\cos 4f+{{x}_{35}}\cos 5f+{{x}_{36}}\cos 6f,\\ 
    {{y}_{3}}=&{{y}_{30}}+{{y}_{31}}\sin f+{{y}_{32}}\sin 2f+{{y}_{33}}\sin 3f+{{y}_{34}}\sin 4f+{{y}_{35}}\sin 5f+{{y}_{36}}\sin 6f,\\ 
    {{z}_{3}}=&{{z}_{30}}+{{z}_{31}}\cos f+{{z}_{32}}\cos 2f+{{z}_{33}}\cos 3f+{{z}_{34}}\cos 4f+{{z}_{35}}\cos 5f+{{z}_{36}}\cos 6f.\\ 
\end{aligned}
\label{jie3}
\end{equation}
During the determination of $(x_{32}, y_{32}, z_{32})$ and $(a_2, b_2)$, a rank-deficiency problem arises. This is a common issue in the process of L-P method. To resolve this problem, we follow the standard treatment adopted in the analytic construction of Halo orbits \citep{richardson1980analytic,jorba1999} and eliminate the indeterminacy by setting $y_{32} = 0$ and $z_{32} = 0$, thereby ensuring the system of equations becomes solvable. 
As a result, the second-order coefficients of correction terms are
\begin{equation}
{{a}_{2}}=\frac{4{{C}_{x2}^{(3)}}-(5+2{{c}_{2}}){{S}_{y2}^{(3)}}}{(5+2{{c}_{2}})\kappa \alpha } , \quad {{b}_{2}}=\frac{1}{\beta }{{C}_{z2}^{(3)}},
\end{equation}
and those third-order non-zero coefficients of coordinates are given by
\begin{equation}
    \begin{aligned}
        & {{x}_{30}}=-\frac{{{C}_{x0}^{(3)}}}{1+2{{c}_{2}}}, \quad {{x}_{31}}=-\frac{(-1+6{{c}_{2}}){{C}_{x1}^{(3)}}+2(5+2{{c}_{2}}){{S}_{y1}^{(3)}}}{6\left( 3+3{{c}_{2}}+2c_{2}^{2} \right)},\\
        &{{x}_{32}}=-\frac{{{C}_{x2}^{(3)}}}{5+2{{c}_{2}}} ,\quad{{x}_{33}}=-\frac{(41+10{{c}_{2}}){{C}_{x3}^{(3)}}-(30+12{{c}_{2}}){{S}_{y3}^{(3)}}}{10\left( 23+11{{c}_{2}}+2c_{2}^{2} \right)},\\
        &{{x}_{34}}=-\frac{(19+6{{c}_{2}}){{C}_{x4}^{(3)}}-2(5+2{{c}_{2}}){{S}_{y4}^{(3)}}}{3{{\left( 9+2{{c}_{2}} \right)}^{2}}},\\ 
         & {{x}_{35}}=-\frac{(121+42{{c}_{2}}){{C}_{x5}^{(3)}}-10(5+2{{c}_{2}}){{S}_{y5}^{(3)}}}{42\left( 63+27{{c}_{2}}+2c_{2}^{2} \right)},\\
         & {{x}_{36}}=-\frac{4(11+4{{c}_{2}}){{C}_{x6}^{(3)}}-3(5+2{{c}_{2}}){{S}_{y6}^{(3)}}}{8\left( 181+76{{c}_{2}}+4c_{2}^{2} \right)},\\
         &{{y}_{31}}=-\frac{(5+2{{c}_{2}}){{C}_{x1}^{(3)}}-(5+7{{c}_{2}}+2c_{2}^{2}){{S}_{y1}^{(3)}}}{3\left( 3+3{{c}_{2}}+2c_{2}^{2} \right)},\\ 
        & {{y}_{33}}=-\frac{-(15+6{{c}_{2}}){{C}_{x3}^{(3)}}+(25+15{{c}_{2}}+2c_{2}^{2}){{S}_{y3}^{(3)}}}{5\left( 23+11{{c}_{2}}+2c_{2}^{2} \right)},\\ 
        & {{y}_{34}}=-\frac{-8(5+2{{c}_{2}}){{C}_{x4}^{(3)}}+(85+44{{c}_{2}}+4c_{2}^{2}){{S}_{y4}^{(3)}}}{12{{\left( 9+2{{c}_{2}} \right)}^{2}}},\\ 
        & {{y}_{35}}=-\frac{-5(5+2{{c}_{2}}){{C}_{x5}^{(3)}}+(65+31{{c}_{2}}+2c_{2}^{2}){{S}_{y5}^{(3)}}}{21\left( 63+27{{c}_{2}}+2c_{2}^{2} \right)},\\ 
        &{{y}_{36}}=-\frac{-12(5+2{{c}_{2}}){{C}_{x6}^{(3)}}+(185+84{{c}_{2}}+4c_{2}^{2}){{S}_{y6}^{(3)}}}{32\left( 181+76{{c}_{2}}+4c_{2}^{2} \right)},\\
        & {{z}_{30}}=-\frac{1}{4}{{C}_{z0}^{(3)}} , \quad  {{z}_{31}}=\frac{1}{3}{{C}_{z1}^{(3)}}, \quad  {{z}_{33}}=-\frac{1}{5}{{C}_{z3}^{(3)}}  , \quad  {{z}_{34}}=-\frac{1}{12}{{C}_{z4}^{(3)}},\\
        &{{z}_{35}}=-\frac{1}{21}{{C}_{z5}^{(3)}}  , \quad {{z}_{36}}=-\frac{1}{32}{{C}_{z6}^{(3)}}.
          \label{A4}
\end{aligned}
\end{equation}
In summary, the third-order expansion of M2N1 ME-Halo orbits can be organized as 
\begin{equation}\label{Eq23}
\begin{aligned}
{{x}}=&{{x}_{20}}+{{x}_{30}}+({{x}_{21}}+{{x}_{31}})\cos f+(\alpha+{{x}_{32}})\cos 2f+({{x}_{23}}+{{x}_{33}})\cos 3f\\
&+({{x}_{24}}+{{x}_{34}})\cos 4f+{{x}_{35}}\cos 5f+{{x}_{36}}\cos 6f,\\ 
{{y}}=& {{y}_{20}} + {{y}_{30}}+({{y}_{21}}+{{y}_{31}})\sin f+\kappa\alpha\sin 2f+({{y}_{23}}+{{y}_{33}})\sin 3f\\
&+({{y}_{24}}+{{y}_{34}})\sin 4f+{{y}_{35}}\sin 5f+{{y}_{36}}\sin 6f,\\ 
{{z}}=&{{z}_{20}}+{{z}_{30}}+({{z}_{21}}+{{z}_{31}})\cos f+\beta\cos 2f+({{z}_{23}}+{{z}_{33}})\cos 3f\\
&+({{z}_{24}}+{{z}_{34}})\cos 4f+{{z}_{35}}\cos 5f+{{z}_{36}}\cos 6f,\\ 
\end{aligned}
\end{equation}
and the correction terms can be expressed as
\begin{equation}\label{Eq24}
\Delta_1 = a_0  + a_2 = 0,\quad
\Delta_2 = b_0  + b_2 = 0,
\end{equation}
where $a_1=b_1=0$ is considered. Considering the explicit expressions of $a_2$ and $b_2$, we can further arrange the correction terms as follows:
\begin{equation}\label{Eq25}
\begin{aligned}
\Delta_1 &= a_0 + a_{200} e^2 + a_{020} \alpha^2 + a_{002} \beta^2 = 0,\\
\Delta_2 &= b_0 + b_{200} e^2 + b_{020} \alpha^2 + b_{002} \beta^2 = 0,
\end{aligned}
\end{equation}
where the coefficients $a_{200}$, $a_{020}$, $a_{002}$, $b_{200}$, $b_{020}$, and $b_{002}$ are provided by Eq. \eqref{A4} in the Appendix. The constraints given by Eq. \eqref{Eq25} provide implicit relationships among $e$, $\alpha$, and $\beta$. In particular, if $e$ is treated as an independent parameter, the amplitudes $\alpha$ and $\beta$ can be uniquely determined by
\begin{equation}
\begin{aligned}
\alpha &= \sqrt{\frac{{b_{002} (a_0 + a_{200} e^2) - a_{002} (b_0 + b_{200} e^2)}}{{-a_{020} b_{002} + a_{002} b_{020}}}},\\
\beta  &= \sqrt{\frac{{-b_{020} (a_0 + a_{200} e^2) + a_{020} (b_0 + b_{200} e^2)}}{{-a_{020} b_{002} + a_{002} b_{020}}}}.
\end{aligned}
\end{equation}

\subsection{High-order analytical solution}
\label{subsec33}

Here we extend to higher-order series expansions of ME-Halo orbits in the ERTBP. Due to the perturbations introduced by nonlinear terms, the general solution of ME-Halo orbits near the collinear libration points in the ERTBP can be expanded as a power series in terms of the eccentricity $e$, the in-plane amplitude $\alpha$, and the out-of-plane amplitude $\beta$ in the following manner:
\begin{equation}
\begin{aligned}
    x&=\sum{x_{ijk}^{l}}{{e}^{i}}{{\alpha }^{j}}{{\beta }^{k}}\cos (lf),  \\
    y&=\sum{y_{ijk}^{l}}{{e}^{i}}{{\alpha }^{j}}{{\beta }^{k}}\sin (lf),  \\
    z&=\sum{z_{ijk}^{l}}{{e}^{i}}{{\alpha }^{j}}{{\beta }^{k}}\cos (lf).  \\
\end{aligned}
\label{xyz}
\end{equation}
At the same time, the correction terms $\Delta_1$ and $\Delta_2$ are expanded as power series in terms of the eccentricity and amplitudes in the following form:
\begin{equation}
    {{\Delta }_{1}}=\sum{{{a}_{ijk}}{{e}^{i}}{{\alpha }^{j}}{{\beta }^{k}}},\quad
    {{\Delta }_{2}}=\sum{{{b}_{ijk}}{{e}^{i}}{{\alpha }^{j}}{{\beta }^{k}}}.
\end{equation}
Here, $i, j, k \in \mathbb{N}$, $l \in \mathbb{Z}$, and $l$ has the same parity as $i$. The unknown coefficients $(x_{ijk}^{l}, y_{ijk}^{l}, z_{ijk}^{l}, a_{ijk}^{l}, b_{ijk}^{l})$ are iteratively determined using the perturbation method. For convenience, we introduce two orders: the order $N_1$ is defined in terms of the orbital eccentricity, and $N_2$ is the order with respect to the amplitudes $\alpha$ or $\beta$. The total order of the analytical solution is $N = N_1 + N_2$. For any combination of indices $i, j, k, l$, the conditions $0 \le i \le N_1$, $0 \le j + k \le N_2$ should be satisfied. For M2N1 ME-Halo orbits, the allowable range of $l$ is $-i - 2(j + k) \le l \le i + 2(j + k)$. Considering the symmetry properties of cosine and sine functions, it is enough to compute the terms with $0 \le l \le i + 2(j + k)$.

As for the linear solution, the associated order is denoted as $(n_1=0, n_2=1)$, corresponding to zeroth-order in eccentricity and first-order in amplitudes. Thus, the coefficients of the first-order solution can be further expressed as 
\begin{equation*}
 x_{010}^{2}=1,\quad y_{010}^{2}=\kappa,\quad z_{001}^{2}=1,\quad a_{000}= -\frac{9+c_2(5-2c_{2})}{5+2{{c}_{2}}},\quad b_{000}= {{c}_{2}}-4.  
\end{equation*}

In the process of constructing higher-order solutions, it is necessary to distinguish the known terms from those unknown terms in the equations of motion. The known terms at order $(n_1, n_2)$ arise from the right-hand side of the equations of motion, excluding the correction terms. All the known terms are collected on the right-hand side and denoted as $X_{ijk}^{l}$, $Y_{ijk}^{l}$, and $Z_{ijk}^{l}$. The unknowns at order $(n_1, n_2)$ include the coordinate coefficients $(x_{ijk}^{l}, y_{ijk}^{l}, z_{ijk}^{l})$ of the same order, and the correction-related coefficients $(a_{ijk}, b_{ijk})$ of order $(n_1, n_2 - 1)$. For convenience, the unknown terms of analytical solution at order $(n_1,n_2)$ are summarized in Table \ref{Tab1}.

\begin{table}
\centering
\caption{Summary of unknown terms for the $n$-th order solution.}
\label{rme}
\begin{tabular*}{0.99\textwidth}{@{\extracolsep{\fill}} ccccc}
\toprule
$\Delta_1 y$ & $\Delta_2 z$& $x,y,z$  & \textbf{$x',y'$ } &  $x'',y'',z''$  \\
\midrule
${{a}_{000}}y_{ijk}^{l}$ & ${{b}_{000}}z_{ijk}^{l}$ &$x_{ijk}^l,y_{ijk}^l,z_{ijk}^l$ &$lx_{ijk}^l,ly_{ijk}^l,lz_{ijk}^l$ &$l^2x_{ijk}^l,l^2y_{ijk}^l,l^2z_{ijk}^l$ \\
${{a}_{ij-1k}}\kappa \delta_{2l} $ & ${{b}_{ijk-1}}\delta_{2l}$ & \quad & \quad & \quad\\
\botrule
\end{tabular*}
\label{Tab1}
\end{table}

Application of perturbation method indicates that those terms of the same order on both sides must be equal, allowing us to formulate the system of linear equations at order $(n_1, n_2)$ as follows:
\begin{equation}
\begin{aligned}
    -({{l}^{2}}+1+2{{c}_{2}})x_{ijk}^{l}-2ly_{ijk}^{l}&=X_{ijk}^{l},\\ 
   -({{l}^{2}}+1-{{c}_{2}}+{{a}_{000}})y_{ijk}^{l}-2lx_{ijk}^{l}-{{a}_{ij- 1k}}\kappa {{\delta }_{2l}}&=Y_{ijk}^{l},\\ 
    (-{{l}^{2}}+{{c}_{2}}-{{b}_{000}})z_{ijk}^{l}-{{b}_{ijk-1}}{{\delta }_{2l}}&=Z_{ijk}^{l},
\end{aligned}
\end{equation}
where $\delta_{ij}$ is the Kronecker symbol, with $\delta_{ij} = 1$ if $i = j$, and $\delta_{ij} = 0$ otherwise. The unknown coefficients including $(x_{ijk}^{l}, y_{ijk}^{l}, z_{ijk}^{l})$ and $(a_{ij-1k}, b_{ijk-1})$ can be determined by solving the linear system of equations in the cases of $l = 2$ and $l \ne 2$. See Table \ref{TableB2} for the coefficients of series expansions up to order 3 about the M2N1 ME-Halo orbits around $L_2$ in the system of $\mu = 0.0001$.

\begin{table}[htbp]
    \centering
    \caption{Coefficients of correction terms and coordinates of the third-order series expansion of M2N1 ME-Halo orbits around $L_2$ in the dynamical model of $\mu = 0.0001$.}
    \label{rme}
    \begin{tabular}{ccccSS}
    \toprule
    \textbf{$i$} & \textbf{$j$} & \textbf{$k$ } & \textbf{$ $}&\multicolumn{1}{c} \textbf{$a_{ijk}$} &\multicolumn{1}{c} \textbf{$b_{ijk}$}  \\
    \midrule
        0 & 0 & 0 & & .815473465266E-01 & -.185347359371E+00 \\
        0 & 0 & 2 & & .806857647185E+00 & -.535380180484E+00 \\
        0 & 2 & 0 & & -.425982982820E+01 & .804332948695E+01 \\
        2 & 0 & 0 & & .897012913088E+00 & .114325067073E+01 \\
    \midrule
    \textbf{$i$} & \textbf{$j$} & \textbf{$k$ } & \textbf{$l$ }& \multicolumn{1}{c}\textbf{$x_{ijk}^l$ or $z_{ijk}^l$} & \multicolumn{1}{c}\textbf{$y_{ijk}^l$} \\
    \midrule
    0 & 0 & 1 & 2 & .100000000000E+01 & \\
    0 & 1 & 0 & 2 & .100000000000E+01 & -.315732632031E+01 \\
    0 & 0 & 2 & 0 & -.255082536862E+00 & \\
    0 & 0 & 2 & 4 & -.111141395671E+00 & .670187851100E-01 \\
    0 & 1 & 1 & 4 & -.366864180416E+00 & \\
    1 & 0 & 1 & 1 & .469108773438E+00 & \\
    1 & 1 & 0 & 3 & .509046867362E-02 & .704151869059E+00 \\
    1 & 1 & 0 & 1 & .790992345009E+00 & -.165102706230E+01 \\
    0 & 0 & 3 & 6 & -.187785890778E-01 & \\
    0 & 1 & 2 & 2 & .370681934798E-01 & \\
    0 & 1 & 2 & 6 & .865444702303E-01 & .169450480988E-01 \\
    0 & 2 & 1 & 6 & .401585394226E+00 & \\
    0 & 3 & 0 & 2 & .110077910069E+01 & \\
    0 & 3 & 0 & 6 & -.788114962028E+00 & -.848345233831E+00 \\
    1 & 0 & 2 & 1 & -.692012517163E-01 & -.798581043538E-01 \\
    1 & 0 & 2 & 3 & -.113912738869E+00 & .940113342615E-01 \\
    1 & 0 & 2 & 5 & .652929569041E-01 & -.335586474518E-01 \\
    1 & 1 & 1 & 1 & .100871431754E+01 & \\
    1 & 1 & 1 & 3 & -.565989773830E+00 & \\
    1 & 1 & 1 & 5 & .187342231102E+00 & \\
    1 & 2 & 0 & 1 & -.461663256351E+00 & -.853250204681E+00 \\
    1 & 2 & 0 & 3 & .883266646720E+00 & .847119291068E+00 \\
    1 & 2 & 0 & 5 & -.487462337750E+00 & -.260887486736E+00 \\
    2 & 0 & 1 & 0 & -.108685090518E-01 & \\
    2 & 0 & 1 & 4 & .916480528772E-01 & \\
    2 & 1 & 0 & 0 & .145496172504E+00 & .372615077613E-01 \\
    2 & 1 & 0 & 2 & -.696658926851E-01 & \\
    2 & 1 & 0 & 4 & -.100029035334E-01 & -.236124768923E+00 \\
    \botrule
    \end{tabular}
    \label{TableB2}
\end{table}

It is worth noting that ME-Halo orbits must satisfy the following correction conditions:
\begin{equation}
    {{\Delta }_{1}}(e,\alpha ,\beta )=0,\quad {{\Delta }_{2}}(e,\alpha ,\beta )=0,
    \label{eqys}
\end{equation}
which indicate that, among the eccentricity $e$, and amplitudes $\alpha$ and $\beta$, only one is independent. Given one of these parameters, the remaining ones can be determined from Eq. \eqref{eqys}. 

\section{Results}
\label{sec4}

The high-order analytical solution of ME-Halo orbits in the ERTBP has been implemented in the previous section. Simulation results are presented in this section, and all results are shown in the synodic coordinate system with origin at the libration point $L_2$.

\subsection{Accuracy analysis}
\label{subsec41}

\begin{figure}[h]
\centering
\includegraphics[width=0.8\textwidth]{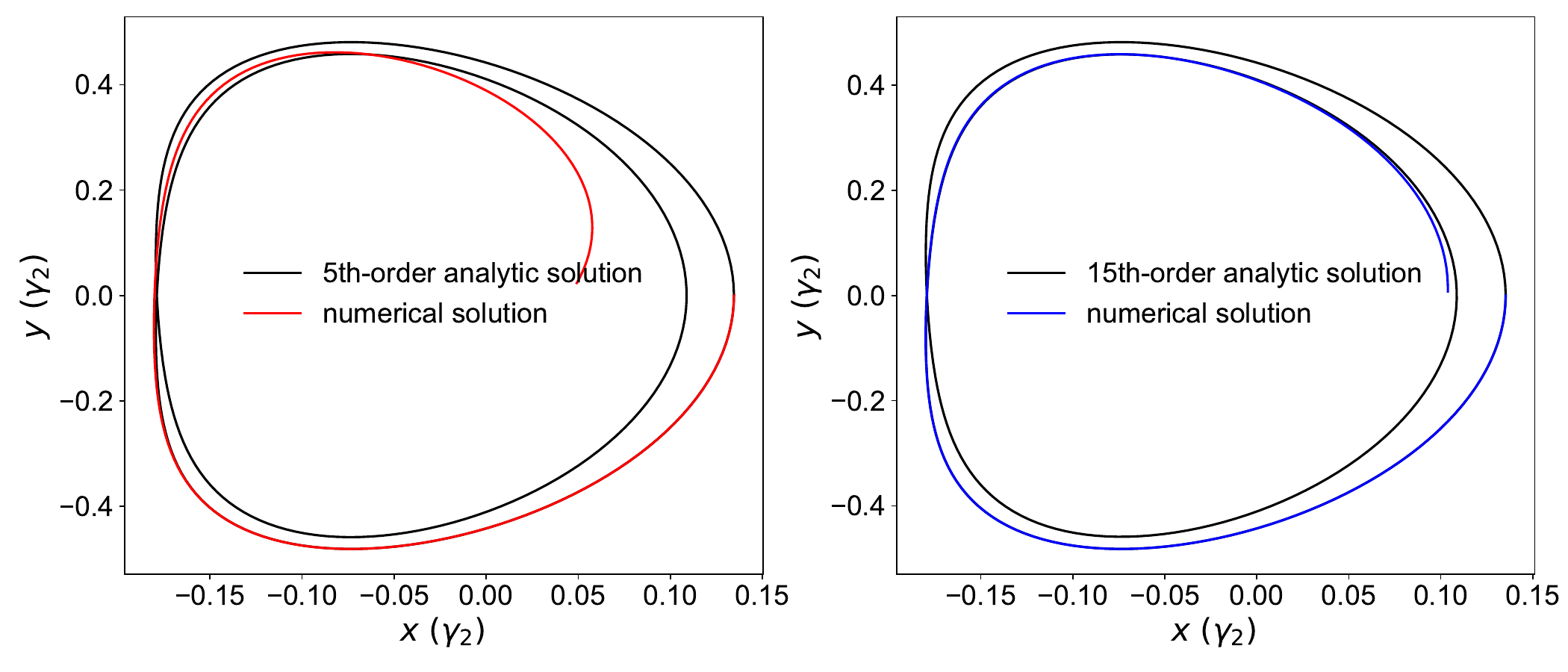}
\includegraphics[width=0.8\textwidth]{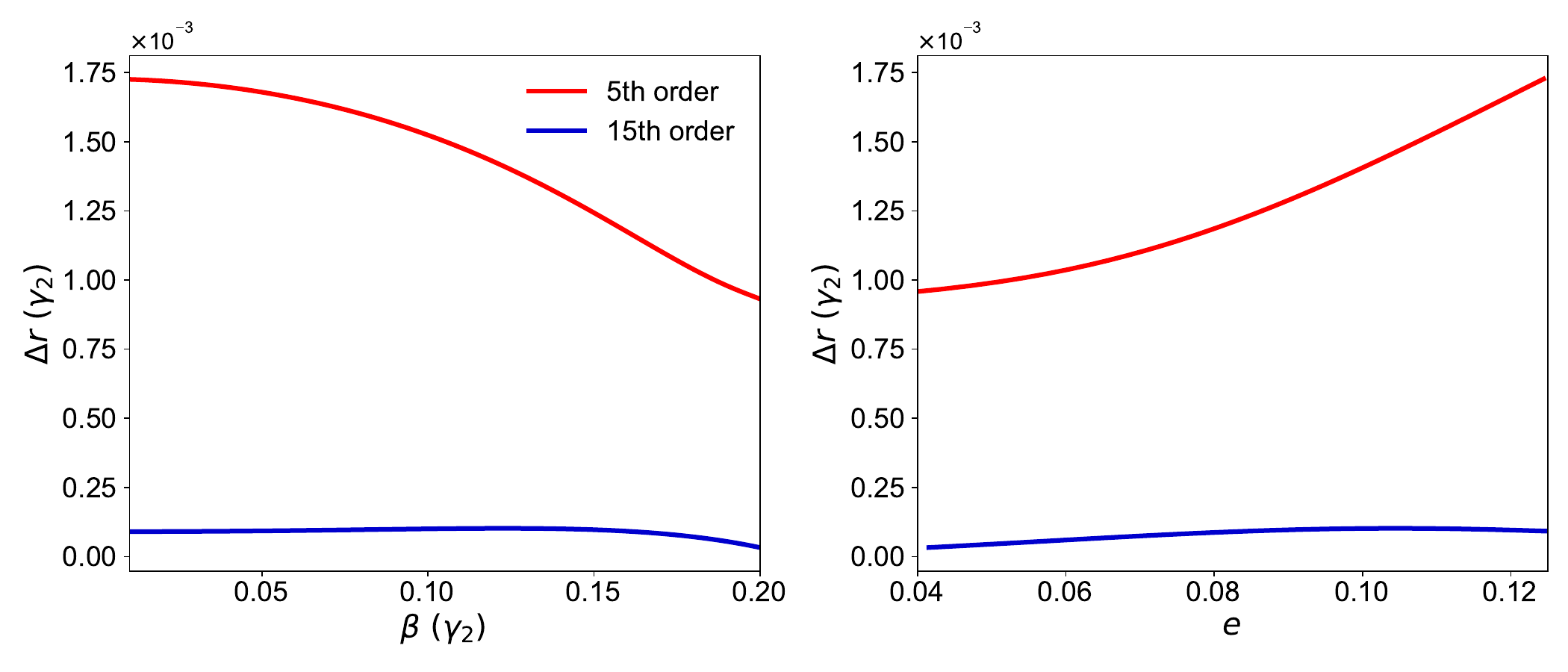}
\caption{Analytical ME-Halo orbits together with the associated numerically propagated orbits (\textit{top panels}), and position deviation between analytical and numerical orbits evaluated at a quarter period (\textit{bottom panels}). Analytical expansions up to order $n = 5$ and $n = 15$ are considered under the dynamical model of $\mu = 0.0001$. In the top panels, the out-of-plane amplitude is fixed at $\beta = 0.1$, and it holds $(e,\alpha)=(0.109232,0.149158)$ for $n = 5$ and $(e,\alpha)=(0.112684,0.149471)$ for $n=15$.}\label{fig5}
\end{figure}

To verify the accuracy of the analytical solutions, the method of numerical integration is employed. Initially, the analytical solution of ME-Halo orbits at different orders are taken to produce initial conditions. Then, the equations of motion are numerically integrated by the RKF78 integrator.
The top panels of Fig. \ref{fig5} shows a comparison between analytical and numerical orbits. It shows that the analytical orbit produced from the 15th-order expansion has a better agreement with the corresponding numerically propagated orbit than the 5th-order version.

To quantitatively assess the accuracy of analytical expansions, the Euclidean norm of the position difference between analytical and numerical orbits is evaluated at a quarter period. The position deviation $\Delta r$ is taken as an index of accuracy \citep{lei2013,lei2014high}. The bottom panels of Fig. \ref{fig5} illustrates how the position deviation $\Delta r$ changes with eccentricity $e$ and the out-of-plane amplitude $\beta$ for the 5th- and 15th-order analytical expansions. It is observed that a higher-order expansion holds a higher-level accuracy. In general, the error decreases with the out-of-plane amplitude $\beta$ and increases with the eccentricity of primaries $e$.

Table \ref{mudr} lists the values of $e,\alpha$ and $\Delta r$ for varying $\mu$, with $\beta$ fixed at 0.04. The results indicate that $\Delta r$ increases monotonically with $\mu$. For systems with $\mu < 0.0001$, the deviation remains on the order of $10^{-5}$, whereas as $\mu$ approaching $0.001$, it grows up to the order of $10^{-3}$. This is because a higher $\mu$ holds a larger ME halo orbit, as shown in Fig. \ref{fig1}.

\begin{table}[htbp]
    \centering
    \caption{Values of $e,\alpha$ and $\Delta r$ with $\beta$ fixed at 0.04. The position deviation $\Delta r$ is evaluated at a quarter period of the corresponding ME-Halo orbit.}
    \label{mudr}
    \begin{tabular*}{0.99\textwidth}{@{\extracolsep{\fill}} cccc}
    \toprule
    \textbf{$\mu$} & \textbf{$e$} & \textbf{$\alpha$ } & \textbf{$\Delta r(\gamma_2)$ } \\
    \midrule
    0.00008 & 0.070892 & 0.146131 & 2.953167E-05\\
    0.0001 & 0.124957& 0.146461 & 9.169962E-05\\
    0.0003 & 0.280288& 0.148482 & 4.751369E-04\\
    0.0005 & 0.338801& 0.149693 & 7.114512E-04\\
    0.001 & 0.414768 & 0.151685 & 1.115248E-03\\
    0.0122 & 0.683231& 0.163858 & 4.888446E-03\\
    \bottomrule
    \end{tabular*}
\end{table}

\subsection{Characteristic curves of ME-Halo orbits}
\label{subsec42}

\begin{figure}[h]
\centering
\includegraphics[width=0.8\textwidth]{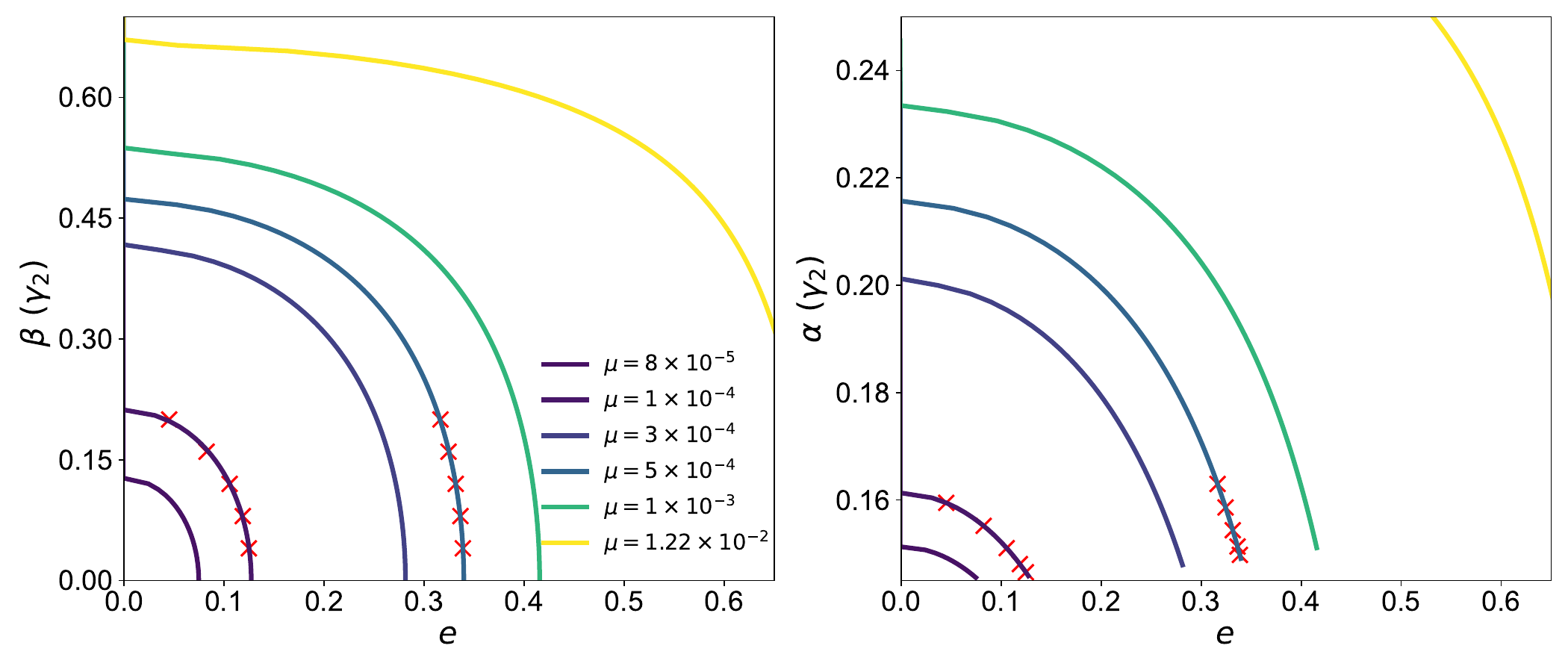}
\caption{Characteristic curves of the M2N1 ME-Halo orbits produced by means of 15th-order series expansions in dynamical models with different $\mu$. The ME-Halo orbits marked by red crosses are to be presented in Figs. \ref{fig2} and \ref{fig3}.}\label{fig1}
\end{figure}

As for analytical solution of ME-Halo orbits, the parameters include the mass ratio $\mu$, the eccentricity $e$, the in-plane amplitude $\alpha$, and the out-of-plane amplitude $\beta$. Fig. \ref{fig1} illustrates the relationship among $e$, $\alpha$, and $\beta$, corresponding to characteristic curves of ME-Halo orbits under different dynamical models specified by $\mu$.It is observed that the mass parameter $\mu$ determines the distribution of characteristic curve of ME-Halo orbits. In particular, a lower value of $\mu$ corresponds to a smaller range of $e$ and $\alpha$ (or $\beta$), while higher value of $\mu$ corresponds to a larger range of $e$ and $\alpha$ (or $\beta$).

\begin{figure}[h]
\centering
\includegraphics[width=0.99\textwidth]{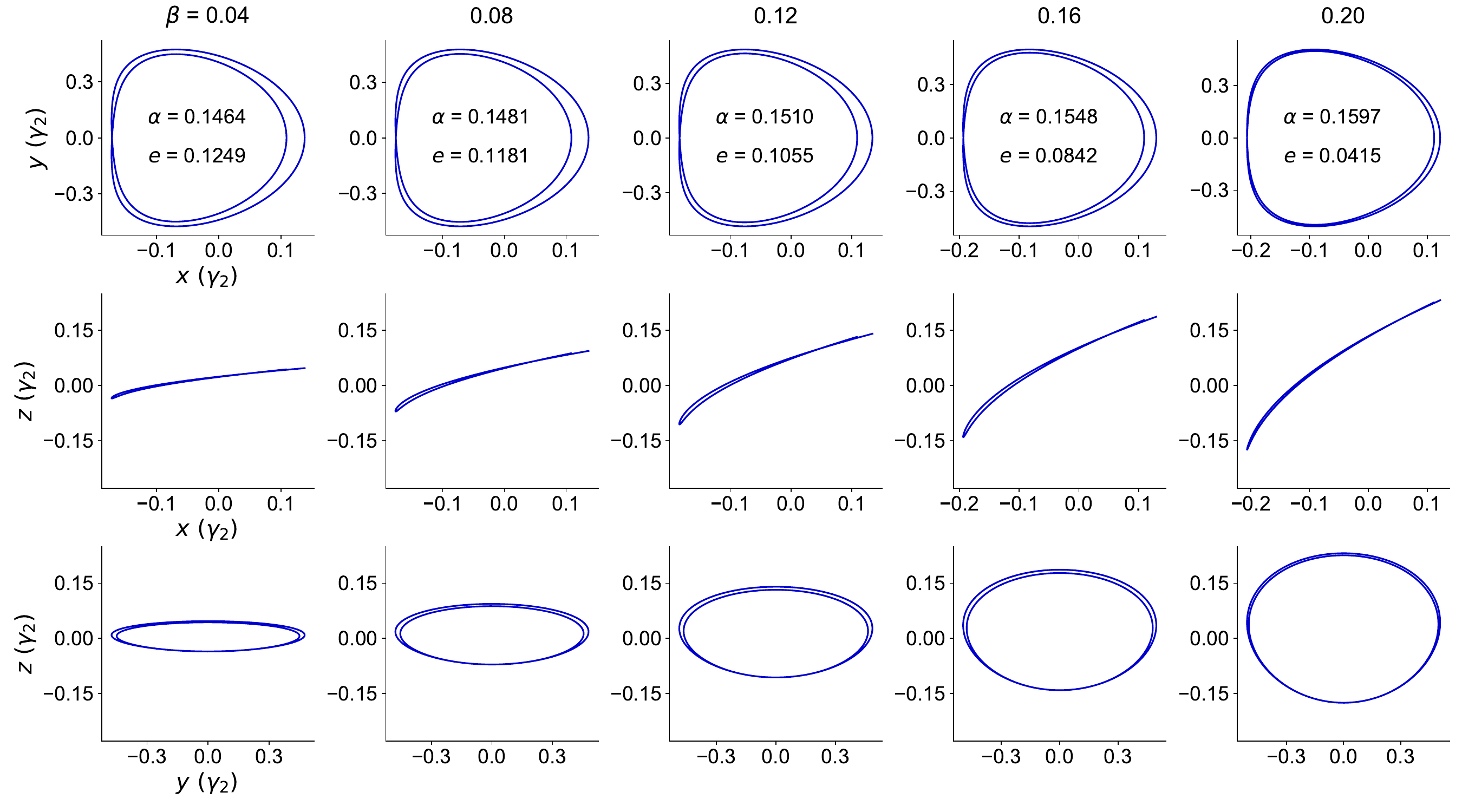}
\caption{Analytical M2N1 ME-Halo orbits in the northern periapsis group, shown in different coordinate planes as $\beta$ changes from 0.04 to 0.2 in the dynamical model of $\mu = 0.0001$. The parameters of ME-Halo orbits, including $e$, $\alpha$ and $\beta$, are provided in the top-row panels (please see Fig. \ref{fig1} for their location on the characteristic curve).}\label{fig2}
\end{figure}

\begin{figure}[h]
\centering
\includegraphics[width=0.99\textwidth]{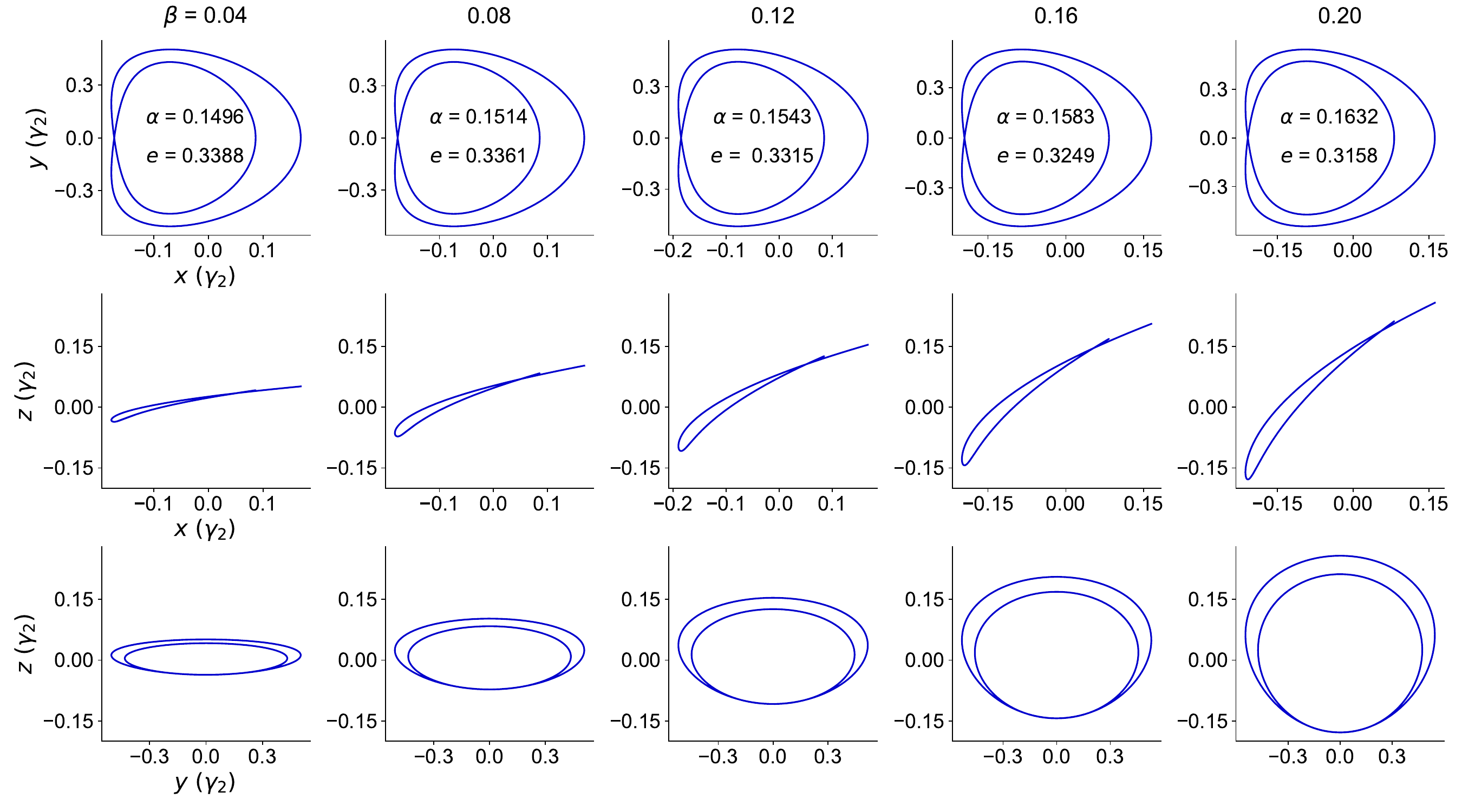}
\caption{Same as Fig. \ref{fig2} but for the dynamical model of $\mu = 0.0005$.}\label{fig3}
\end{figure}

Figs. \ref{fig2} and \ref{fig3} present the projections of analytical M2N1 ME-Halo orbits to coordinate planes for the ERTBPs of $\mu = 0.0001$ and $\mu = 0.0005$, as the out-of-plane amplitude $\beta$ varies from 0.04 to 0.2 with step of 0.04. Their locations on the characteristic curves are marked by red crosses in Fig. \ref{fig1}. As $\beta$ increases (from the left to the right), the amplitude of motion in the $z$ direction gradually increases, causing the orbits projected onto the $x$-$z$ and $y$-$z$ planes to expand. On the other hand, the eccentricity $e$ decreases with $\beta$, which results in the two loops of the orbits projected onto the $x$–$y$ and $y$-$z$ planes moving closer together.

The Halo orbit is symmetric with respect to the $xy$-plane, thus there exist two families of solutions: the northern family with $z > 0$ and the southern family with $z < 0$. Based on their initial positions, ME-Halo orbits can also be categorized into the periapsis group, which starts from the periapsis of the primary (i.e., $f_0 = 0$), and into the apoapsis group, which starts from the apoapsis (i.e., $f_0 = \pi$). Figs. \ref{fig2} and \ref{fig3} show ME-Halo orbits inside the northern periapsis group. The corresponding orbits inside the remaining three groups are shown in Figs. \ref{FIGA3}-\ref{FIGA5} in Appendix \ref{secA3}.

\section{Numerically corrected orbits}
\label{sec5}

A fixed-time single shooting method is taken to identify numerically corrected ME-Halo orbits, with the analytical solution as initial guesses.

\subsection{Differential correction}\label{subsec51}

For completeness, a brief introduction to the single shooting method is provided \citep{howell1984}. In the barycentric synodic frame, the state vector of Halo orbit is denoted by $\boldsymbol{X} = (X, Y, Z, {X}', {Y}', {Z}')^T$. Owing to the orbit’s symmetry about the $xz$ plane and the fact that its velocity is perpendicular to this plane, the initial state is chosen at the intersection point. Accordingly, the analytical solution provides the initial condition at $\boldsymbol{X_0}= (X_0, 0, Z_0, 0, {Y}'_0, 0)^T$, with the set of adjustable parameters denoted by $\boldsymbol{Q}=(X_0, Z_0, {Y}'_0)^T$.

Due to the symmetry, Halo orbits will cross the $xz$ plane again after half a period. Therefore, it is only necessary to compute the trajectory over half a period and then propagate it forward by the remaining half to obtain the complete periodic orbit. To ensure the condition of periodicity, the state at the half period must satisfy the following constraint, denoted by
\begin{equation}
    \boldsymbol{F}(\boldsymbol{Q})=(Y_{T/2},{X}'_{T/2},{Z}'_{T/2})^T=\boldsymbol{0}.
    \label{zqys}
\end{equation}
The Newton iteration method is employed to solve Eq. \eqref{zqys}, and the correction process stops when the assumed tolerance is reached.

\subsection{Numerically corrected orbits in the ERTBP}
\label{subsec52}

The 15th-order analytical solution is used to produce the analytical orbit, which serves as the initial guess. Then, the orbit is numerically refined to yield a high-accuracy periodic orbit based on the correction procedure.

\begin{figure}[h]
\centering
\includegraphics[width=0.99\textwidth]{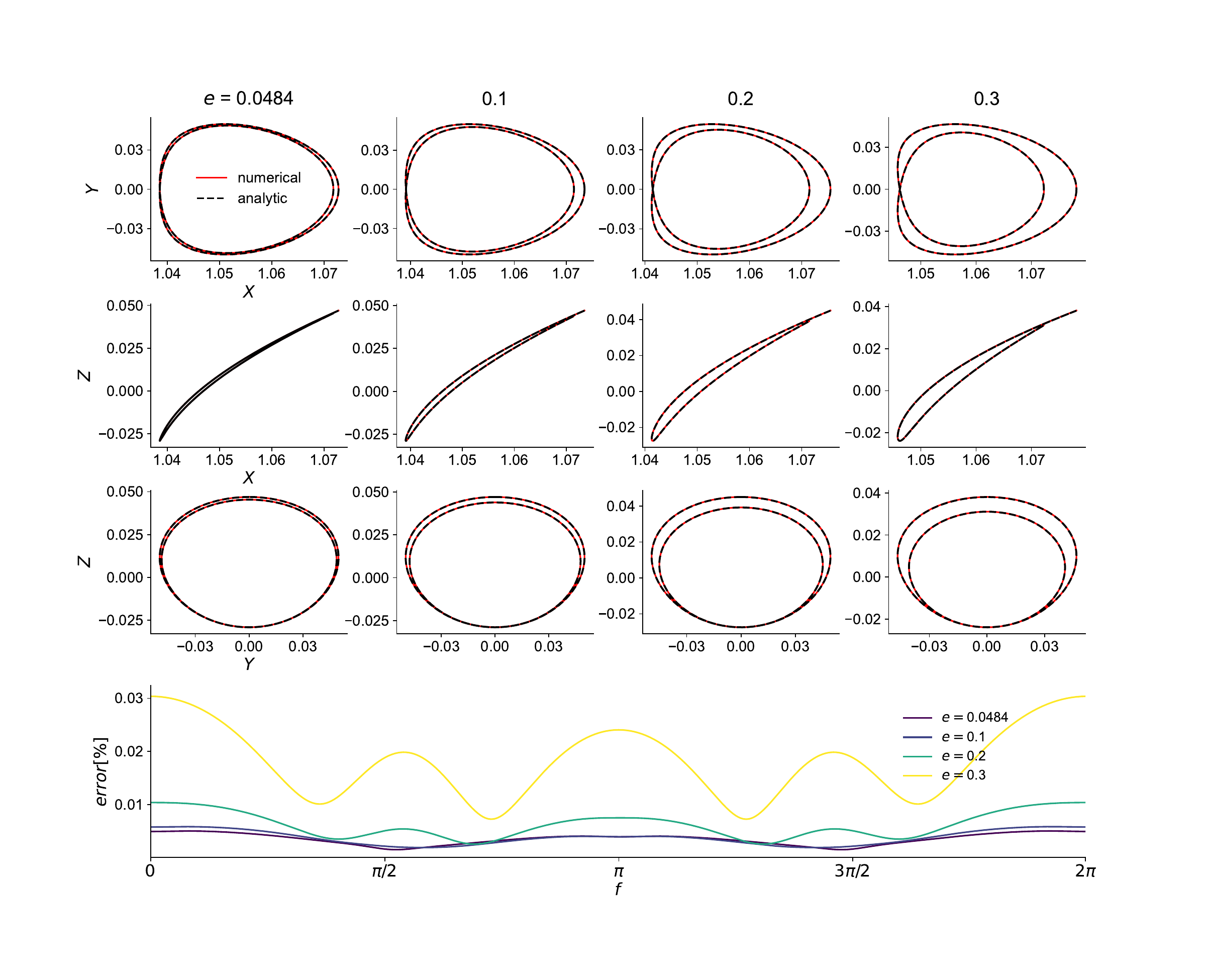}
\caption{The numerically corrected orbits together with the analytical M2N1 ME-Halo orbits (\textit{panels in the top three rows}), and the normalized error between the analytical and numerical orbits as a function of the true anomaly (\textit{bottom-row panels}) in the Sun--Jupiter systems with different eccentricities.}\label{fig6}
\end{figure}

\begin{figure}[h]
\centering
\includegraphics[width=0.99\textwidth]{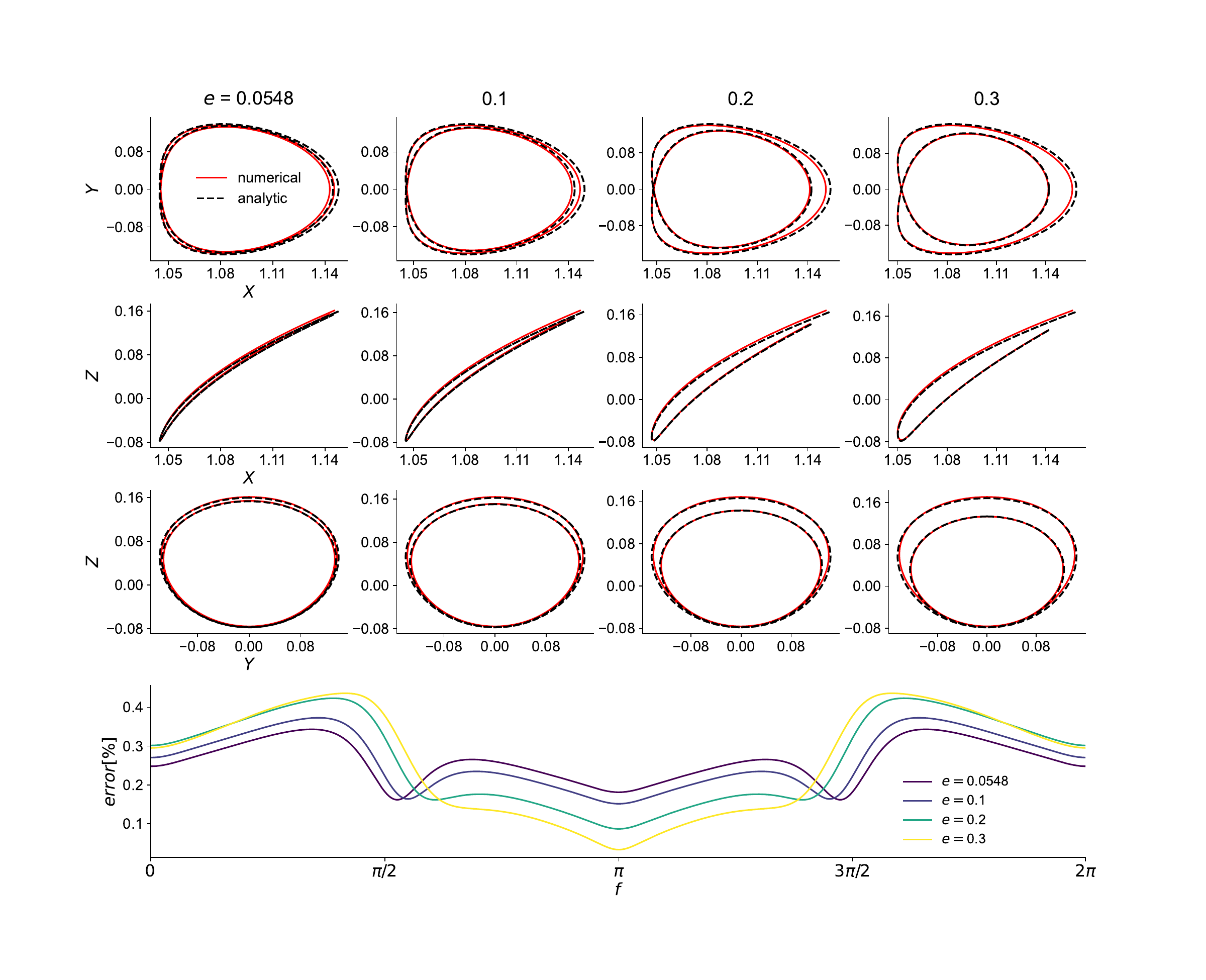}
\caption{Same as \ref{fig6} but for the Earth--Moon system.}\label{fig7}
\end{figure}

To assess the applicability of the analytical solution with varying parameters, the mass ratio $\mu$ is fixed for the Sun–Jupiter and Earth–Moon systems, while the orbital eccentricity is changed. Table \ref{table2} lists the parameters $(e,\alpha,\beta)$ of the considered representative ME-Halo orbits. Analytical orbits are then taken as initial guesses for the single shooting method. The panels in the first-three rows of Figs. \ref{fig6} show the M2N1 ME-Halo orbits around the $L_2$ point in the Sun--Jupiter, including the analytical orbits and the numerically corrected orbits. Fig. \ref{fig7} corresponds to the case of Earth--Moon ERTBPs. In these plots, the black dashed lines represent the analytical orbits, and the red solid lines show the numerically corrected M2N1 ME-Halo orbits. The unit of length is the instantaneous distance between the primaries. It is observed that there is a good agreement between analytical orbits and numerically corrected orbits.

Additionally, a comparison is also made for ME-Halo orbits before and after correction in terms of the normalized error, defined by \citep{Asano2015Approximating}
\begin{equation}
\mathrm{error}(f) = \frac{\lVert \boldsymbol{X}_{\mathrm{ana}}(f) - \boldsymbol{X}_{\mathrm{num}}(f) \rVert}{\lVert \boldsymbol{X}_{\mathrm{num}}(f) \rVert},
\label{eq:normalized_error}
\end{equation}
where $\boldsymbol{X}_{\mathrm{ana}}(f)$ and $\boldsymbol{X}_{\mathrm{num}}(f)$ are the analytical and numerical orbits, respectively. Please refer to the bottom panels of Figs. \ref{fig6} and \ref{fig7} for the curves of $\mathrm{error}(f)$. It is observed that even in cases with relatively large eccentricity, the constructed analytical solutions remain high-accuracy approximations for periodic orbits in the full dynamical model. For $\mu=0.00095$, the normalized error remains below 0.05\% over the entire range of true anomaly. Similarly, for $\mu=0.0122$, the error is also bounded within 0.5\% across all values of $f$.

\begin{table}[htbp]
    \centering
    \caption{Parameters of M2N1 ME-Halo orbits under the Sun--Jupiter and Earth--Moon models but with different primary eccentricities.}
    \label{rme}
    \begin{tabular*}{0.99\textwidth}{@{\extracolsep{\fill}} cccc}
    \toprule
    \textbf{$\mu$} & \textbf{$e$} & \textbf{$\alpha$ } & \textbf{$\beta$ } \\
    \midrule
        \multirow{4}{*}{0.00095 (S--J)} & 0.048400 & 0.231182 & 0.526408 \\
         & 0.100000 & 0.229257 & 0.518509 \\
         & 0.200000 & 0.220807 & 0.483350 \\
         & 0.300000 & 0.202339 & 0.402660 \\
        \hline
        \multirow{4}{*}{0.01220 (E--M)} & 0.054800 & 0.286478 & 0.664588 \\
         & 0.100000 & 0.285941 & 0.662558 \\
         & 0.200000 & 0.283497 & 0.653372 \\
         & 0.300000 & 0.278859 & 0.636102 \\
    \botrule
    \end{tabular*}
    \label{table2}
\end{table}

The numerical results confirm the effectiveness of the analytical solution developed in this study under the framework of ERTBP, fully demonstrating its reliability in terms of both accuracy and applicability. This provides a robust theoretical foundation for its use in trajectory design for deep space exploration missions. In particular, ME-Halo orbits, serving as three-dimensional parking and transfer orbits, offer significant advantages, including low energy consumption, short orbital periods, and high transfer flexibility. These characteristics make them increasingly valuable orbital assets for missions targeting the Earth-Moon system and beyond.

\section{Conclusion}\label{sec6}

In this study, we propose a novel method for constructing high-order expansions of ME-Halo orbits near collinear libration points in the framework of ERTBP. Specifically, two correction terms are introduced into the equations of motion. The first correction term, denoted by $\Delta_1 y$, is employed to enforce a specific commensurability between the forced frequency ($n$) and either the in-plane frequency ($\omega$) or the out-of-plane frequency ($\nu$). The second correction term, denoted by $\Delta_2 z$, is introduced to achieve the frequency degeneracy condition of $\omega = \nu$ (also known as the Halo orbit condition).

The perturbation method is employed to construct high-order expansions of ME-Halo orbits. In this approach, both the coordinate variables ($x$, $y$, $z$) and the correction terms ($\Delta_1$, $\Delta_2$) are expressed as power series in three parameters: the orbital eccentricity of the primary body ($e$), the in-plane amplitude ($\alpha$), and the out-of-plane amplitude ($\beta$). For ME-Halo orbits, these parameters ($e$, $\alpha$, $\beta$) must strictly satisfy the correction conditions $\Delta_1(e, \alpha, \beta) = 0$ and $\Delta_2(e, \alpha, \beta) = 0$, implying that only one of them can be treated as an independent variable.

Based on the constructed analytical solution, characteristic curves of ME-Halo orbits can be derived analytically in the $(e, \alpha)$ and $(e, \beta)$ spaces under dynamical models with different mass parameters $\mu$. Due to the system’s inherent symmetry, analytical ME-Halo orbits can be systematically categorized into four distinct groups: the southern periapsis group, the southern apoapsis group, the northern periapsis group, and the northern apoapsis group. These families of periodic orbits provide rich insights into the dynamical behavior near collinear libration points in the ERTBP.

The accuracy of the high-order expansions of ME-Halo orbits is evaluated by comparing the analytical solutions with numerically integrated orbits. The results demonstrate that higher-order expansions yield significantly improved precision. Consequently, the analytical solutions developed in this study can serve as highly accurate initial guesses for generating numerically corrected ME-Halo orbits. This capability is validated through the successful construction of numerically corrected ME-Halo orbits in both the Sun–Jupiter and Earth–Moon systems with a large range of orbital eccentricities.  

Future work could further leverage the proposed analytical solution as a high-accuracy initial guess within multiple shooting methods, embedded in a high-fidelity dynamical model constructed using JPL ephemerides. By incorporating perturbative effects such as solar radiation pressure and gravitational influences from additional planetary bodies, the long-term evolution and maintenance requirements of the orbits can be systematically evaluated under realistic dynamical environments. This would enhance the practical robustness and applicability of multi-revolution targeting strategies for deep space missions.

This study focuses exclusively on M2N1 ME-Halo orbits. However, more complex resonant configurations, such as M3N1 and M5N2, are also commonly found in realistic celestial systems \citep{peng2015}. These configurations exhibit richer dynamical behaviors and hold significant research interest and promising applications. Future work could extend the current methodology to encompass a broader class of resonant structures, with the goal of developing a unified high-order analytical framework capable of addressing stronger nonlinearities and more intricate resonance conditions. Such a framework would offer a more general and robust theoretical foundation for the analysis and trajectory design of ME-Halo orbits within three-body gravitational systems.

\backmatter

\bmhead{Acknowledgements}
The authors wish to thank the anonymous reviewer for insightful comments, which help to improve the readability of the paper. This work is financially supported by the National Natural Science Foundation of China (Nos. 12573063 and 12233003) and the China Manned Space Program with grant no. CMS-CSST-2025-A16.

\bmhead{Author Contributions}
Xiaoyan Leng performed the research and wrote the main manuscript. Hanlun Lei supervised the research and reviewed the manuscript.

\bmhead{Data availability} 
No datasets were generated or analyzed during the current study.

\section*{Declarations}
\bmhead{Conflict of interest}
The authors declare no conflict of interest.

\begin{appendices}
\section{Coefficients of the equations of motion}\label{secA1}

The coefficients of the equations of motion at the second order are
\begin{equation}\label{A1}
    \begin{aligned}
        & {{C}_{x0}^{(2)}}=\frac{3}{4}{{c}_{3}}\left( 2{{\alpha }^{2}}-{{\beta }^{2}}-{{\alpha }^{2}}{{\kappa }^{2}} \right) ,\quad {{C}_{x1}^{(2)}}=-\frac{1}{2}\left( 1+2{{c}_{2}} \right)e\alpha,\\
        & {{C}_{x3}^{(2)}}=-\frac{1}{2}\left( 1+2{{c}_{2}} \right)e\alpha ,\quad {{C}_{x4}^{(2)}}=\frac{3}{4}{{c}_{3}}\left( 2{{\alpha }^{2}}-{{\beta }^{2}}+{{\alpha }^{2}}{{\kappa }^{2}} \right),\\
        & {{S}_{y1}^{(2)}}=\frac{1}{2}\kappa \left( -1+{{c}_{2}} \right)e\alpha ,\quad {{S}_{y3}^{(2)}}=\frac{1}{2}\kappa \left( -1+{{c}_{2}} \right)e\alpha  ,\quad {{S}_{y4}^{(2)}}=-\frac{3}{2}{{c}_{3}}\kappa {{\alpha }^{2}},\\
        & {{C}_{z0}^{(2)}}=-\frac{3}{2}{{c}_{3}}\alpha \beta ,\quad {{C}_{z1}^{(2)}}=\frac{1}{2}\left( -1+{{c}_{2}} \right)e\beta ,\\
        &{{C}_{z3}^{(2)}}=\frac{1}{2}\left( -1+{{c}_{2}} \right)e\beta  ,\quad {{C}_{z4}^{(2)}}=-\frac{3}{2}{{c}_{3}}\alpha \beta,
    \end{aligned}
\end{equation}
and the coefficients of the equations of motion at the third order are
\begin{equation}\label{A2}
\begin{aligned}
    & {{C}_{x0}^{(3)}}=\frac{1}{4}[ -2e{{x}_{21}}\left( 1+2{{c}_{2}} \right)+{{e}^{2}}\alpha \left( 1+2{{c}_{2}} \right) ],\\ 
    & {{C}_{x1}^{(3)}}=-\frac{1}{4}[ 6{{c}_{3}}\left( -2{{x}_{21}}\alpha -2{{x}_{23}}\alpha +{{z}_{21}}\beta +{{z}_{23}}\beta +{{y}_{21}}\alpha \kappa +{{y}_{23}}\alpha \kappa  \right),\\
    & +\left( 4+8{{c}_{2}} \right)e{{x}_{20}}-3e{{c}_{3}}\left( {{\beta }^{2}}-2{{\alpha }^{2}}+{{\kappa }^{2}}{{\alpha }^{2}} \right)],\\ 
    & {{C}_{x2}^{(3)}}=\frac{1}{2}[ -e\left( 1+2{{c}_{2}} \right)\left( {{x}_{21}}+{{x}_{23}} \right)+3{{c}_{3}}\left( 4{{x}_{20}}\alpha +2{{x}_{24}}\alpha -2{{z}_{20}}\beta -{{z}_{24}}\beta -{{y}_{24}}\alpha \kappa  \right)\\
    &+{{e}^{2}}\alpha \left( 1+2{{c}_{2}} \right)-3{c}_{4}\alpha \left( 3{\beta }^{2}+{\alpha }^{2}{\kappa }^{2}-2{\alpha }^{2} \right) ],\\ 
    & {{C}_{x3}^{(3)}}=\frac{1}{8}[ 12{{c}_{3}}\left( 2{{x}_{21}}\alpha -{{z}_{21}}\beta +{{y}_{21}}\alpha \kappa  \right) \\ 
    & -e\left( 4{{x}_{24}}+8{{c}_{2}}{{x}_{24}}+6{{c}_{3}}{{\alpha }^{2}}-3{{c}_{3}}{{\beta }^{2}}+3{{c}_{3}}{{\alpha }^{2}}{{\kappa }^{2}} \right) ],\\ 
    & {{C}_{x4}^{(3)}}=\frac{1}{4}\left( -2e{{x}_{23}}+{{e}^{2}}\alpha  \right)\left( 1+2{{c}_{2}} \right),\\ 
    & {{C}_{x5}^{(3)}}=\frac{1}{8}[ 12{{c}_{3}}\left( 2{{x}_{23}}\alpha -{{z}_{23}}\beta +{{y}_{23}}\alpha \kappa  \right)-\left( 4+8{{c}_{2}} \right)e{{x}_{24}} -3e{{c}_{3}}\left( -{{\beta }^{2}}+2{{\alpha }^{2}}+{{\kappa }^{2}}{{\alpha }^{2}} \right) ],\\ 
    & {{C}_{x6}^{(3)}}=\frac{1}{2}[ -3{{c}_{4}}\alpha {{\beta }^{2}}+3{{c}_{3}}\left( 2{{x}_{24}}\alpha -{{z}_{24}}\beta +{{y}_{24}}\alpha \kappa  \right)+{{c}_{4}}{{\alpha }^{3}}\left( 2+3{{\kappa }^{2}} \right) ],\\
    & {{S}_{y1}^{(3)}}=\frac{1}{2}[ \left( -1+{{c}_{2}} \right)e{{y}_{22}}+3{{c}_{3}}\alpha \left( {{y}_{21}}-{{y}_{23}}-{{x}_{21}}\kappa +{{x}_{23}}\kappa  \right) ],\\ 
    & {{S}_{y2}^{(3)}}=\frac{1}{8}[ 4e\left( -1+{{c}_{2}} \right)\left( {{y}_{21}}+{{y}_{23}} \right)+8{{a}_{2}}\alpha \kappa -12\alpha {{c}_{3}}\left( {{y}_{24}}+2{{x}_{20}}\kappa -{{x}_{24}}\kappa  \right)\\
    &-4{{e}^{2}}\alpha \kappa \left( -1+{{c}_{2}} \right)+3\alpha {{c}_{4}}\kappa \left( {{\beta }^{2}}-4{{\alpha }^{2}}+3{{\kappa }^{2}}{{\alpha }^{2}} \right) ],\\ 
    & {{S}_{y3}^{(3)}}=\frac{1}{4}[ -6{{c}_{3}}\alpha \left( {{y}_{21}}+{{x}_{21}}\kappa  \right)+2e{{y}_{24}}\left( -1+{{c}_{2}} \right)+3{{c}_{3}}e{{\alpha }^{2}}\kappa ],\\ 
    & {{S}_{y4}^{(3)}}=\frac{1}{4}\left( -1+{{c}_{2}} \right)\left( 2e{{y}_{23}}-{{e}^{2}}\alpha \kappa  \right),\\ 
    & {{S}_{y5}^{(3)}}=\frac{1}{4}[ 2e{{y}_{24}}\left( -1+{{c}_{2}} \right)+3{{c}_{3}}e{{\alpha }^{2}}\kappa -6{{c}_{3}}\alpha \left( {{y}_{23}}+{{x}_{23}}\kappa  \right) ],\\ 
    & {{S}_{y6}^{(3)}}=-\frac{3}{8}\alpha[ 4{{c}_{3}}\left( {{y}_{24}}+{{x}_{24}}\kappa  \right)+{{c}_{4}}\kappa \left( -{{\beta }^{2}}+4{{\alpha }^{2}}+{{\alpha }^{2}}{{\kappa }^{2}} \right)],
 \end{aligned}
 \end{equation}
and
\begin{equation}\label{A3}
    \begin{aligned}
    & {{C}_{z0}^{(3)}}=\frac{1}{4}\left[ 2e{{z}_{21}}\left( -1+{{c}_{2}} \right)+{{e}^{2}}\beta \left( 1-{{c}_{2}} \right) \right],\\ 
     & {{C}_{z1}^{(3)}}=\left( -1+{{c}_{2}} \right)e{{z}_{20}}+\frac{3}{2}{{c}_{3}}e\alpha \beta -\frac{3}{2}{{c}_{3}}\left( {{z}_{21}}\alpha +{{z}_{23}}\alpha +{{x}_{21}}\beta +{{x}_{23}}\beta  \right),\\ 
    & {{C}_{z2}^{(3)}}=\frac{1}{8}\left[ 4e\left( -1+{{c}_{2}} \right)\left( {{z}_{21}}+{{z}_{23}} \right)-4{{e}^{2}}\beta \left( -1+{{c}_{2}} \right)\right.\\
     &\left.-12{{c}_{3}}\left( 2\alpha {{z}_{20}}+\alpha {{z}_{24}}+2\beta {{x}_{20}}+\beta {{x}_{24}} \right)+\beta \left( 8{{b}_{2}}+9{{c}_{4}}{{\beta }^{2}}-36{{c}_{4}}{{\alpha }^{2}}+3{{c}_{4}}{{\alpha }^{2}}{{\kappa }^{2}} \right) \right],\\ 
      & {{C}_{z3}^{(3)}}=\frac{1}{4}\left[ 2e{{z}_{24}}\left( -1+{{c}_{2}} \right)+3{{c}_{3}}e\alpha \beta -6{{c}_{3}}\left( {{z}_{21}}\alpha +{{x}_{21}}\beta  \right) \right],\\ 
    & {{C}_{z4}^{(3)}}=\frac{1}{4}\left( 2e{{z}_{23}}-{{e}^{2}}\beta  \right)\left( 1-{{c}_{2}} \right), \\ 
    &{{C}_{z5}^{(3)}}=\frac{1}{4}\left[ 2\left( -1+{{c}_{2}} \right)e{{z}_{24}}+3{{c}_{3}}e\alpha \beta -6{{c}_{3}}\left( {{z}_{23}}\alpha +{{x}_{23}}\beta  \right)\right],\\
    &{{C}_{z6}^{(3)}}=-\frac{3}{8}\left[ -{{c}_{4}}{{\beta }^{3}}+4{{c}_{3}}\left( {{z}_{24}}\alpha +{{x}_{24}}\beta  \right)+{{c}_{4}}{{\alpha }^{2}}\beta \left( 4+{{\kappa }^{2}} \right) \right]. 
    \end{aligned}
\end{equation}
The coefficients of correction terms associated with the third-order analytical solution are given by
\begin{equation}\label{A4}
    \begin{aligned}
        a_{200}&=\frac{1}{30 (5 + 2 c_2)^2 (3 + 3 c_2 + 2 c_2^2) (23 + 11 c_2 + 2 c_2^2)}[-54594 - 104625 c_2 - 39721 c_2^2\\
        &+ 28869 c_2^3 + 29103 c_2^4 + 8768 c_2^5 + 1272 c_2^6 + 48 c_2^7 - 16 c_2^8],\\
        a_{020}&=\frac{3}{128(1 + 2 c_2) (5 + 2 c_2)^2 (9 + 2 c_2)^2}[c_3^2 (73945 + 29900 c_2 + 245916 c_2^2\\
        & + 208032 c_2^3 + 65904 c_2^4  + 9408 c_2^5 + 576 c_2^6)  -  c_4(9 + 2 c_2)^2 (723 + 1886 c_2 \\
        & + 2168 c_2^2 + 3056 c_2^3 + 1008 c_2^4 + 96 c_2^5) ],\\
        a_{002}&=-\frac{3}{8 (1 + 2 c_2) (5 + 2 c_2)^2 (9 + 2 c_2)^2}[  c_3^2 (7325 + 1480 c_2 - 2120 c_2^2 - 736 c_2^3 - 48 c_2^4)\\
        &+ (9 + 2 c_2)^2 c_4 (-167 - 314 c_2 + 44 c_2^2 + 8 c_2^3) ],\\
        b_{200}&= -\frac{1}{30}(16 - 17 c_2 + c_2^2),\\
        b_{020}& = \frac{3}{128(1 + 2 c_2) (9 + 2 c_2)^2}[ c_3^2(-7325 - 1480 c_2 + 2120 c_2^2 + 736 c_2^3 + 48 c_2^4)\\
        &- (9 + 2 c_2)^2 c_4(-167 - 314 c_2 + 44 c_2^2 + 8 c_2^3)  ],\\
        b_{002} &= \frac{3}{8}  \left[\frac{(505 + 260 c_2 + 36 c_2^2) c_3^2}{(1 + 2 c_2) (9 + 2 c_2)^2} - 3 c_4 \right].
    \end{aligned}
\end{equation}

\clearpage
\section{Multi-revolution Halo orbits in different groups}\label{secA3}

\begin{figure}[h]
\centering
\includegraphics[width=0.99\textwidth]{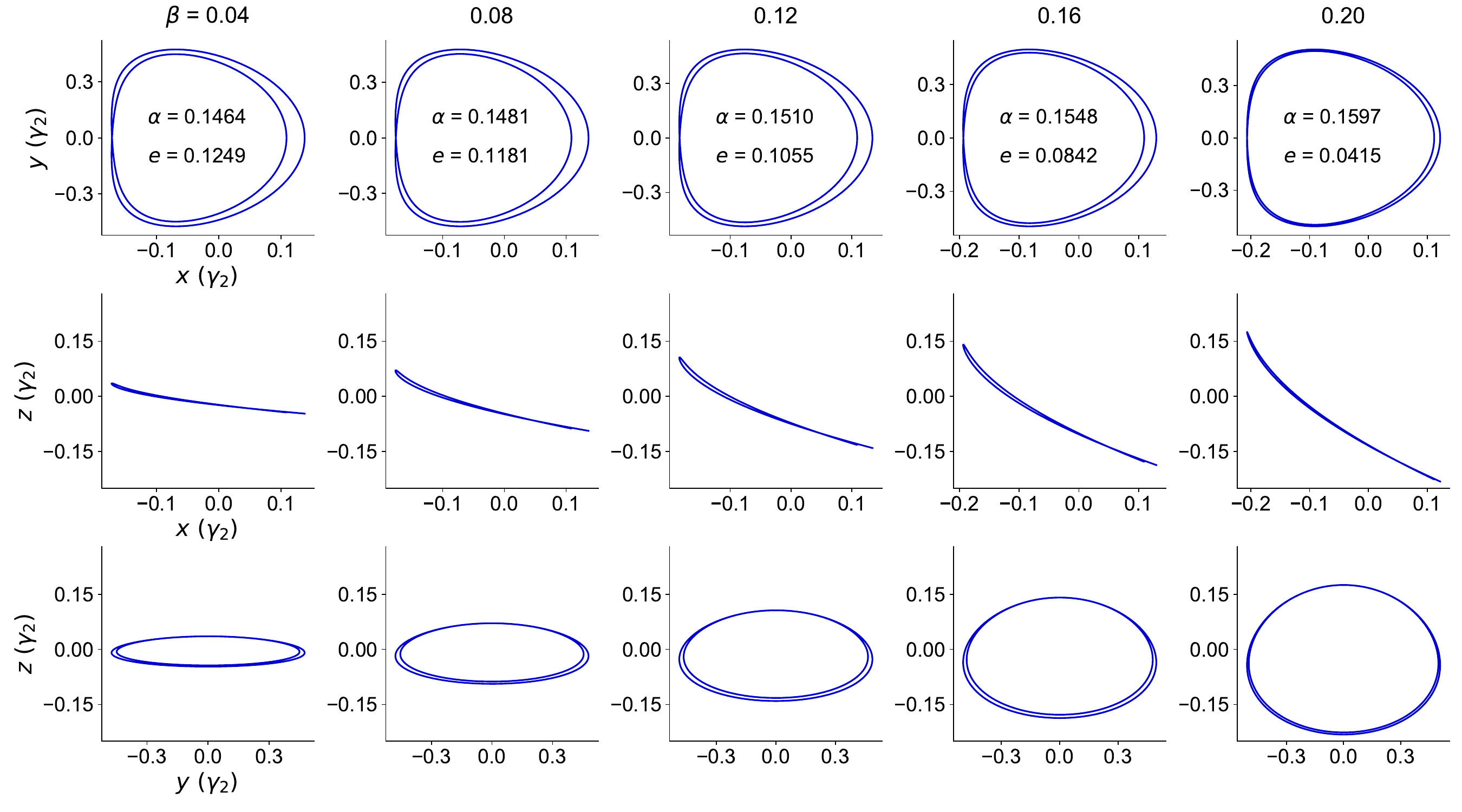}
\caption{Analytical M2N1 ME-Halo orbits in the southern periapsis group, shown in different coordinate planes as $\beta$ varies from 0.04 to 0.2 in the dynamical model of $\mu = 0.0001$.}\label{FIGA3}
\end{figure}

\begin{figure}[h]
\centering
\includegraphics[width=0.99\textwidth]{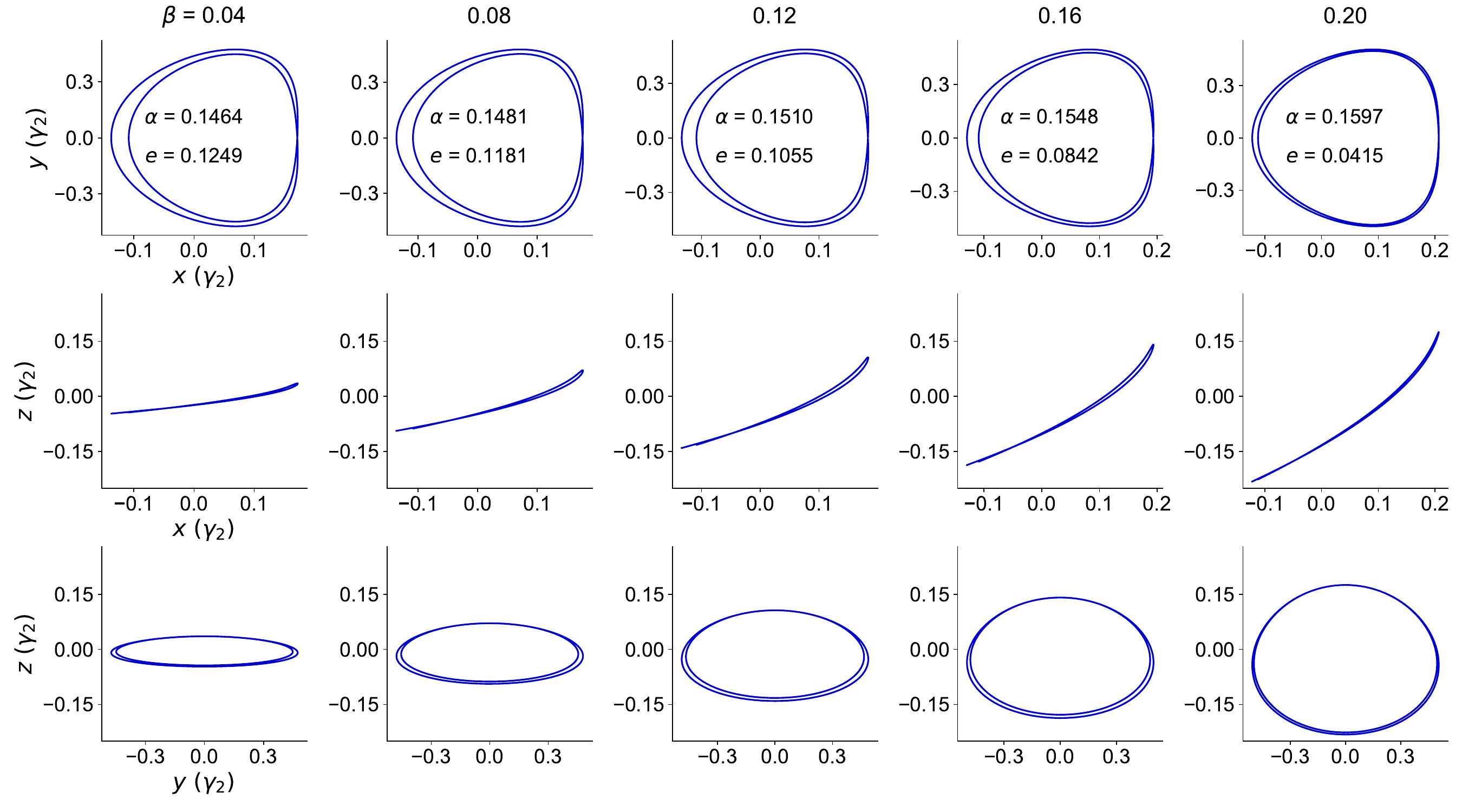}
\caption{Analytical M2N1 ME-Halo orbits in the northern apoapsis group, shown in different coordinate planes as $\beta$ varies from 0.04 to 0.2 in the dynamical model of $\mu = 0.0001$.}\label{FIGA4}
\end{figure}

\begin{figure}[h]
\centering
\includegraphics[width=0.99\textwidth]{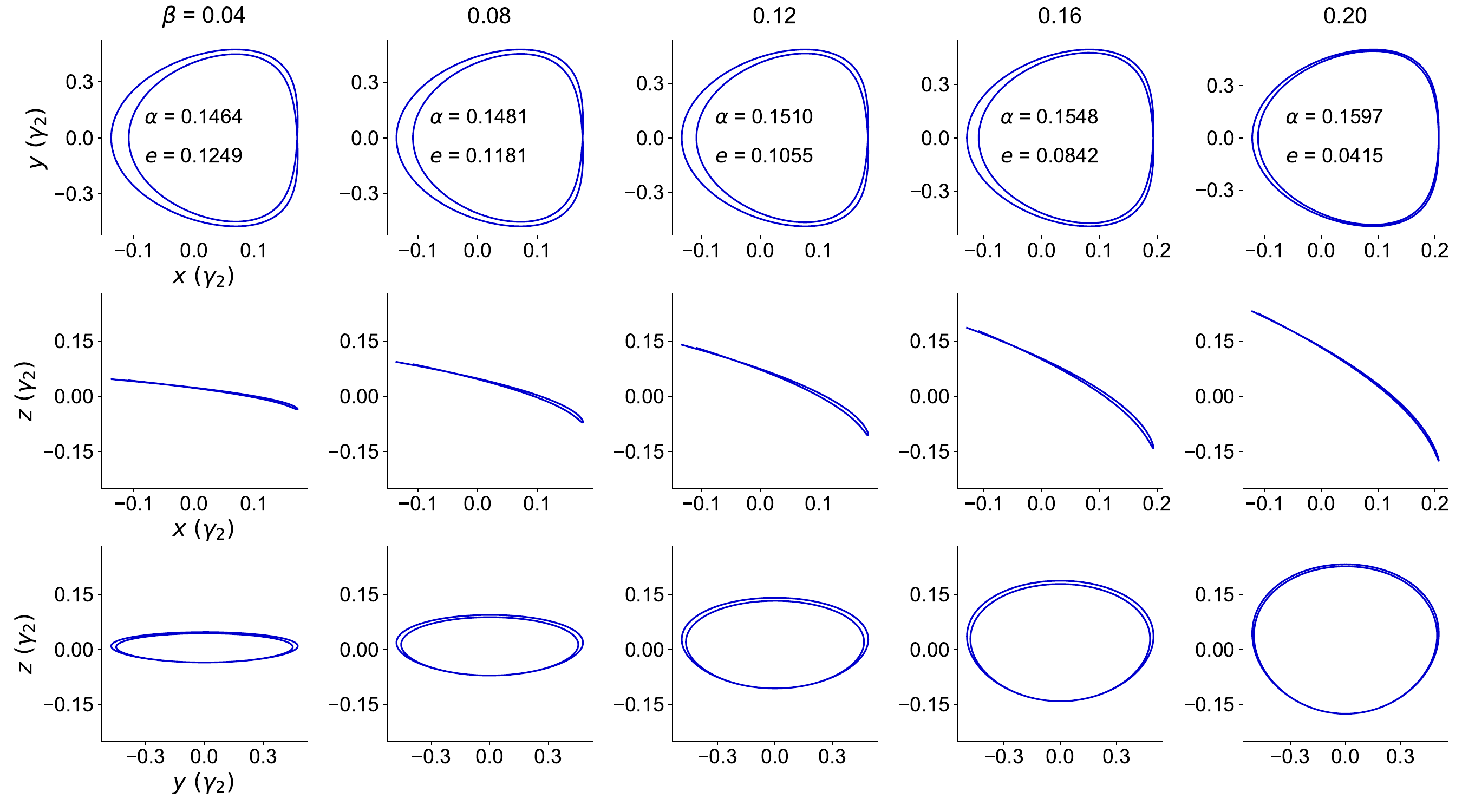}
\caption{Analytical M2N1 ME-Halo orbits in the southern apoapsis group, shown in different coordinate planes as $\beta$ varies from 0.04 to 0.2 in the dynamical model of $\mu = 0.0001$.}\label{FIGA5}
\end{figure}

\end{appendices}

\bibliography{reference}

\end{document}